\PassOptionsToPackage{table,xcdraw}{xcolor}
\documentclass[sigconf, proceedings, natbib=false]{acmart}
\AtBeginDocument{%
  }

\copyrightyear{2026}
\acmYear{2026}
\setcopyright{cc}
\setcctype{by}
\acmConference[PASC '26]{Platform for Advanced Scientific Computing Conference}{June 29-July 01, 2026}{Bern, Switzerland}
\acmBooktitle{Platform for Advanced Scientific Computing Conference (PASC '26), June 29-July 01, 2026, Bern, Switzerland}
\acmDOI{10.1145/3815572.3815738}
\acmISBN{979-8-4007-2734-4/2026/06}

\usepackage{booktabs}
\usepackage{tabularx}
\usepackage{siunitx}
\usepackage{multirow}
\usepackage{booktabs}
\usepackage{tabularx}
\usepackage{pifont}
\newcolumntype{Y}{>{\raggedright\arraybackslash}X}

\usepackage{pgfplots}
\pgfplotsset{compat=1.18}
\usepackage{tikz}
\usetikzlibrary{shadows,positioning,fit,arrows.meta,backgrounds}

\usepackage{booktabs,tabularx,makecell,multirow,ragged2e,array,graphicx}
\usepackage{url}

\definecolor{RowGray}{gray}{0.96} 
\usepackage[most]{tcolorbox}   

\usepackage{numprint} 
\usepackage{placeins}

\usepackage[
  backend=biber,
  style=acmnumeric,
  maxbibnames=3,
  minbibnames=3,
  maxcitenames=2,
  mincitenames=1,
  giveninits=true,
  uniquename=init
]{biblatex}
\begin{document}


\title{astroCAMP: A Community Benchmark and Co-Design Framework for Sustainable SKA-Scale Radio Imaging}

\author{Denisa-Andreea Constantinescu}
\email{denisa.constantinescu@epfl.ch}
\orcid{0000-0001-6736-5715}
\affiliation{%
  \institution{ESL, EPFL}
  \city{Lausanne}
  \country{Switzerland}
}

\author{Rubén~Rodríguez~Álvarez}
\email{ruben.rodriguezalvarez@epfl.ch}
\orcid{0009-0003-1989-1923}
\affiliation{%
  \institution{ESL, EPFL}
  \city{Lausanne}
  \country{Switzerland}
}

\author{Jacques~Morin}
\email{jacques.morin@insa-rennes.fr}
 \orcid{0009-0000-8384-2518}
\affiliation{%
  \institution{Univ Rennes, INSA Rennes, CNRS, IETR - UMR 6164}
  \city{F-35000 Rennes}
  \country{France}
}

\author{Etienne Orliac}
\email{etienne.orliac@epfl.ch}
\orcid{0009-0006-9590-8979}
\affiliation{%
  \institution{SCITAS, EPFL}
  \city{Lausanne}
  \country{Switzerland}
}

\author{Mickaël Dardaillon}
\email{mickael.dardaillon@insa-rennes.fr}
\orcid{0000-0001-6862-2090}
\affiliation{%
  \institution{Univ Rennes, INSA Rennes, CNRS, IETR - UMR 6164}
  \city{F-35000 Rennes}
  \country{France}
}

\author{Sunrise Wang}
\email{sunrise.wang@oca.eu}
\orcid{0000-0002-5038-9531}
\affiliation{%
  \institution{Univ C\^ote d’Azur, OCA, CNRS, J-L.Lagrange - UMR 7293}
  \city{F-06000 Nice}
  \country{France}
}

\author{Hugo Miomandre}
\email{hugo.miomandre@insa-rennes.fr}
\orcid{0009-0005-4832-3292}
\affiliation{%
  \institution{Univ Rennes, INSA Rennes, CNRS, IETR - UMR 6164}
  \city{F-35000 Rennes}
  \country{France}
}

\author{Miguel~Peón-Quirós}
\email{miguel.peon@epfl.ch}
\orcid{0000-0002-5760-090X}
\affiliation{%
  \institution{EcoCloud, EPFL}
  \city{Lausanne}
  \country{Switzerland}
}

\author{Jean-François Nezan}
\email{jean-francois.nezan@insa-rennes.fr}
 \orcid{0000-0002-0609-4592}
\affiliation{%
  \institution{Univ Rennes, INSA Rennes, CNRS, IETR - UMR 6164}
  \city{F-35000 Rennes}
  \country{France}
}

\author{David~Atienza}
\email{david.atienza@epfl.ch}
\orcid{0000-0001-9536-4947}
\affiliation{%
  \institution{ESL, EPFL}
  \city{Lausanne}
  \country{Switzerland}
}

\renewcommand{\shortauthors}{Constantinescu et al.}

\begin{abstract}

The Square Kilometre Array (SKA) will operate one of the world’s largest continuous scientific data systems, sustaining petascale imaging under strict power envelopes. Yet current radio-interfe\-rom\-e\-tric pipelines typically achieve only 4–14\% of hardware peak because of memory and I/O bottlenecks, resulting in high energy, operational, and carbon costs. Progress is further constrained by the absence of standardised cross-layer metrics and survey-level fidelity tolerances for principled hardware–software co-design.
We present \textbf{astroCAMP}, a reproducible benchmarking and co-design framework for SKA-scale imaging. astroCAMP contributes: (1) a unified metric suite spanning performance, utilisation, memory/data-movement behavior, sustainability, economics, and scientific fidelity; (2) standardised SKA-representative datasets, reference outputs, and benchmark configurations for reproducible cross-platform evaluation; (3) a multi-objective co-design formulation linking quality constraints to time-, energy-, carbon-, and cost-to-solution; and (4) a reproducible design-space exploration workflow to derive Pareto-optimal operating regions.
We release datasets, scripts, benchmark results, and a reproducibility kit, and evaluate WSClean+IDG on an AMD EPYC 9334 CPU and an NVIDIA H100 GPU. The evaluation shows substantial end-to-end orchestration and synchronization bottlenecks despite efficient kernels in active phases, limited CPU strong scaling, and location-dependent carbon/cost efficiency under realistic grid and electricity-price assumptions. We further illustrate the use of astroCAMP for heterogeneous CPU–FPGA design-space exploration, and its potential to facilitate the identification of Pareto-optimal operating points for SKA-scale imaging deployments. Lastly, we call on the SKA community to define quantifiable fidelity metrics and thresholds to accelerate principled optimisation for SKA-scale imaging.

\end{abstract}

\ccsdesc[500]{Computer systems organization~Heterogeneous (hybrid) systems}
\ccsdesc[500]{Hardware~Hardware-software codesign}
\ccsdesc[300]{Applied computing~Astronomy}
\ccsdesc[100]{Hardware~Power estimation and optimization}

\keywords{Square Kilometre Array, radio-interferometric imaging, hardware--software co-design, energy- and carbon-aware HPC, reproducible benchmarking}

\maketitle


\section{Introduction}
\label{sec:intro}

\begin{figure}[htbp]
    \centering
    \includegraphics[width=\linewidth]{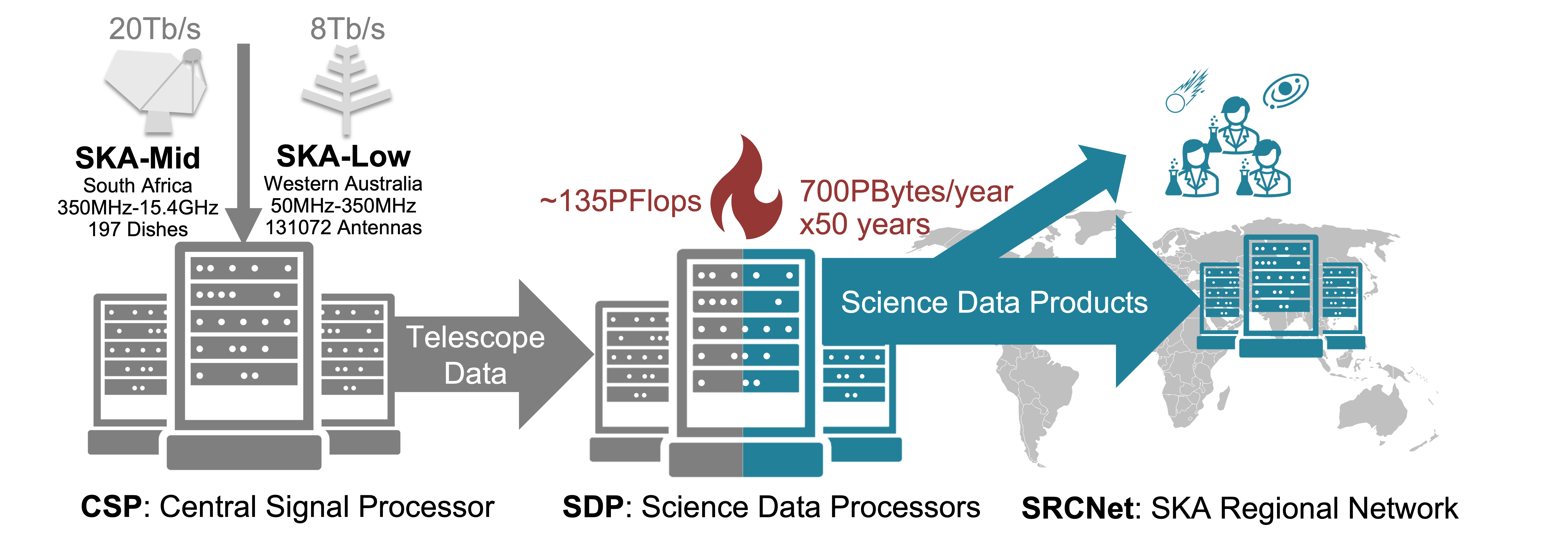}
    \caption{SKA’s infrastructure includes 2 Central Signal Processors (CSPs),
    2 Science Data Processors (SDPs), and a global network of data centers
    (SRCNet). The SKA-Low and SKA-Mid telescopes will stream 28\,Tb/s of data to
    the CSPs, and the SDPs will process it into science data products.}
    \Description{Diagram of the SKA infrastructure showing SKA-Low and SKA-Mid telescopes streaming visibility data to two CSPs and two SDPs, which then feed a distributed SRCNet of regional data centers.}
    \label{fig:ska}
\end{figure}

Modern radio interferometers are entering a regime where
\emph{computing, not photon collection, limits scientific capability}.
Pathfinder and precursor radio interferometers---the Low-Frequency Array (LOFAR), the Murchison Widefield Array (MWA), the Australian Square Kilometre Array Pathfinder (ASKAP), and the Very Large Array (VLA)---already generate visibility, calibration, and imaging workloads that strain current high-performance computing (HPC) facilities, and these pressures intensify for the Square Kilometre Array (SKA), the most data-intensive radio observatory ever built. SKA-Low and SKA-Mid---illustrated in Fig.~\ref{fig:ska}---will stream
$\sim$8--20\,Tb/s of correlated visibilities to two Science Data Processors
(SDPs), each expected to operate within a site-level power cap of 1\,MW for several decades~\cite{dewdney2009ska,broekema2015ska}.
Because both SDPs are located on carbon-intensive grids in South Africa and
Australia, energy efficiency directly determines operational cost,
scientific throughput, and long-term sustainability.

\textbf{Efficiency requirements for SKA SDPs.}
Public SKA design documents~\cite{dewdney2009ska,broekema2015ska}
indicate that each SDP must sustain approximately 16--42 petaFLOP/s (PFLOP/s, $10^{15}$ floating-point operations per second) for SKA-Low and 20--72~PFLOP/s for SKA-Mid, depending on observing mode\footnote{Online SKA SDP performance-model references: \url{https://developer.skao.int/projects/sdp-par-model/en/latest/} and \url{https://ska-telescope.gitlab.io/sdp/ska-sdp-par-model/notebooks/SKA1_Imaging_Performance_Model.html}.}.
For a sustained
compute requirement $C_{\rm sus}$ and power envelope $P_{\rm cap}$, the
facility-level efficiency is
$
\eta_{\rm facility} = \frac{C_{\rm sus}}{P_{\rm cap}}.
$
Only a fraction of this power is available to compute nodes, as modern
HPC systems incur 30--40\% overheads for networking, memory, storage,
and cooling (Power Usage Effectiveness, PUE, of 1.3--1.4). Adopting $f_{\rm overhead}\in[0.3,0.4]$, the
compute-node requirement becomes
$
\eta_{\rm compute} = 
\frac{C_{\rm sus}}{P_{\rm cap}(1 - f_{\rm overhead})}
.
$

\begin{figure}[!htbp]
  \centering
  \includegraphics[width=0.8\linewidth]{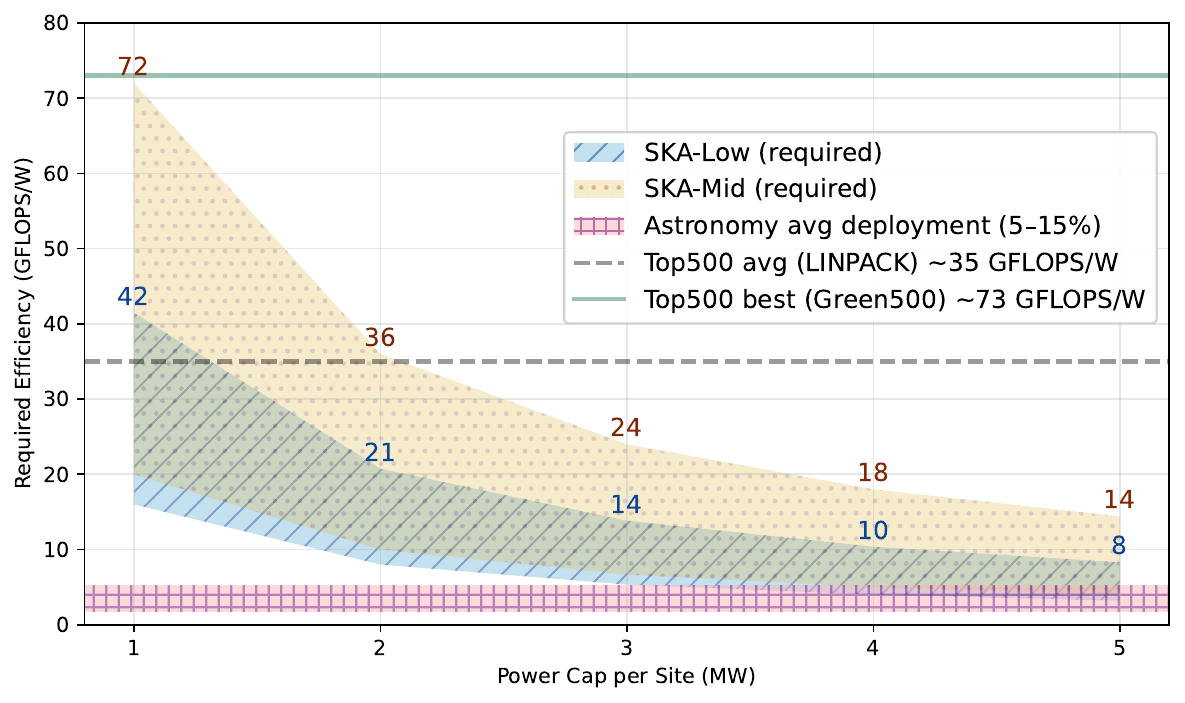}
  \caption{Required energy efficiency for SKA-Low (blue hatched) and SKA-Mid (orange dotted) SDPs vs.\ site power cap. Reference horizontals: \textit{Green500} best (green solid), Top500 average (gray dashed). Pink crosshatched band: astronomy-average application-level deployment.}
  \Description{Plot of required energy efficiency in GFLOP per second per watt for SKA-Low (blue hatched band) and SKA-Mid (orange dotted band) across 1 to 5 megawatt site power caps. A green solid horizontal line marks Green500 best-in-class around 73 GFLOPS/W, a gray dashed horizontal line marks the Top500 average around 35 GFLOPS/W, and a pink crosshatched band near the bottom marks the 5 to 15 percent astronomy-average deployment regime.}
  \label{fig:ska-efficiency-bands}
\end{figure}

Fig.~\ref{fig:ska-efficiency-bands} summarizes the required
efficiency ranges for SKA-Low and SKA-Mid under
1--5~MW power caps. A 2~MW allocation for SKA-Low and SKA-Mid
requires tens of GigaFLOP/s per watt (GFLOP/s/W), comparable to the average
\textit{Green500} systems in November 2025. These typically deliver 30--40~GFLOP/s/W. KAIROS, the
current \textit{Green500} leader, reaches 73.28~GFLOP/s/W~\cite{green5002025nov}. This is comparable to the most computationally demanding operation case for SKA-Low, under a power cap of 1~MW.
However, Green500 benchmarking numbers reflect idealised,
compute-bound kernels measured with LINPACK (the Top500 dense linear-algebra benchmark), whereas SKA imaging workloads---dominated by non-uniform 2D and 3D Fast Fourier Transforms (FFTs)---are in theory memory-limited and in practice I/O-limited.
Despite major algorithmic advances such as
$w$-projection~\cite{cornwell2008noncoplanar},
$w$-stacking~\cite{pratley2019fast},
faceting, and Image Domain Gridding (IDG)~\cite{van2018image}, radio imaging remains dominated by
memory bandwidth (FFTs), data movement, and irregular access patterns.

\textbf{The utilisation gap for imaging software deployments.}
On modern CPU and GPUs,
imaging pipelines typically sustain only 4--14\,\% of peak floating-point performance
\cite{veenboer2020radio}, revealing a fundamental mismatch between imaging
workloads and commodity architectures, as also indicated by the pink band in Fig.~\ref{fig:ska-efficiency-bands}. Earlier modelling~\cite{broekema2015ska} reported only 10–20\% efficiency for FFT- and gridding-dominated workloads, implying that \emph{naïvely scaling} current pipelines to SKA data rates would require tens of megawatts, well beyond SDP site caps. Without major improvements in software efficiency and hardware utilisation, meeting SKA throughput would require substantial over-provisioning and likely exceed the power budget. The core issue is that inefficient implementations can draw significant power even when compute units are under-utilised, producing high operational and carbon costs for limited scientific return.

Closing this gap requires more than incremental optimization of hardware and software tuning. It demands holistic
co-design across algorithms, data structures, and hardware:
\begin{enumerate}
    \item \textbf{Algorithmic optimisation:} reduce data movement through
    hierarchical memory reuse, structured sparsity, and compressed visibility
    formats.
    \item \textbf{Domain-specific accelerators:} exploit GPUs, FPGAs, custom
    architectures, and ASICs optimised for imaging hotspots.
    \item \textbf{Energy-aware orchestration:} use dynamic power scaling and
    locality-aware scheduling to match data-access patterns.
\end{enumerate}

Principled co-design further requires standardised datasets, cross-layer metrics, science-driven image-quality tolerances, and reproducible evaluation methods under representative SKA workloads. Without such framework, optimisation risks becoming a blind pursuit of efficiency gains, obscuring the central question for SKA-scale computing:
\emph{how much scientific fidelity can be traded for performance, energy, and carbon savings without compromising discovery?}

Machine learning, for example,
overcame a similar lack of standardisation through community benchmarks such as
the MLPerf machine-learning benchmark suite~\cite{reddi2020mlperf,farrell2021mlperf}, unified datasets,
accuracy thresholds, and submission protocols.
In large-scale computational physics, mixed-precision solvers exploit application-level convergence tolerances to improve performance while preserving required scientific accuracy~\cite{clark2010mixed}. In climate modeling, the Coupled Model Intercomparison Project (CMIP) defines standardized experimental protocols and evaluation criteria to ensure comparability across models~\cite{eyring2016cmip6}.
Radio astronomy lacks an equivalent effort. These gaps hinder agile innovation in the domain and prevent researchers and developers from comparing figures of merit without re-running large, expensive experiments.

\textbf{Scope and contributions of this work.}
Towards bridging these gaps, this paper introduces \textbf{astroCAMP}, a reproducible benchmarking and co-design framework for radio-interferometric imaging. 
astroCAMP provides the metrics, datasets, and benchmark cases required to evaluate efficiency, sustainability, and quality trade-offs across high-performance and heterogeneous architectures. The contributions are:

\begin{enumerate}

\item \textbf{A unified cross-layer metric suite.}
We define and implement a set of 12+ measurable metrics spanning performance,
energy, carbon, system behaviour, economic cost, and scientific fidelity,
providing a consistent basis for comparing imaging pipelines across CPUs, GPUs,
FPGAs, ASICs, and emerging accelerators.

\item \textbf{Standardised SKA benchmark suite and datasets.}
We release open SKA-Low visibility datasets and reference dirty images on Zenodo as \texttt{astroCAMP-data-v1.0}~\cite{astroCAMPdata2025}, together with parameterised benchmark configurations and reproducibility scripts in the companion astroCAMP GitHub repository~\cite{astroCAMP2025}. This enables reproducible, cross-platform evaluation of SKA imaging pipelines and systematic exploration of hardware--algorithm co-design trade-offs.

\item \textbf{A multi-objective co-design formulation.}  
We formalise imaging as an optimisation problem over algorithmic and architectural parameters, with the objective to minimize time- energy- and carbon-to-solution under explicit quality, cost, and power constraints.

\item \textbf{A reproducible design-space exploration workflow.}  
Using the PREESM framework \cite{PREESM2017}, we demonstrate for a subset of astroCAMP’s metrics suite how astroCAMP can enable multi-objective design-space exploration and the computation of Pareto fronts in the metrics domain space. 

\end{enumerate}

Together, the four contributions form the foundation of the first end-to-end methodology for rigorously evaluating and co-designing SKA-scale imaging pipelines with carbon efficiency as first-class objective. 

The remainder of this paper is organized as follows. Section 2 outlines the sustainability goals and challenges of the SKA SDP. Section~3 presents the astroCAMP framework and co-design methodology. Section 4 introduces the multi-objective co-design formulation and Section 5 the benchmark suite. Section 6 reports the experimental evaluation and results, while Section 7 discusses their implications for hardware–software co-design. Section 8 concludes the paper.

\section{SKA SDP Sustainability Challenges and Gaps}

Why does optimizing efficiency matter?
Recent studies highlight the ecological impact of large-scale scientific computing in astronomy~\cite{portegies2020ecological,Aujoux_2021,dos2024assessment,knodlseder2022estimate}. In response, the SKA Observatory has embedded \textbf{sustainability and net-zero objectives} into its 50-year roadmap, which is aligned with the \emph{UN Sustainable Development Goals}, and adopted the CO$_2$ Performance Ladder to support its net-zero transition~\cite{skaoSustainability2024,skaoEnvFootprint2024}.

For the SDPs, the challenge is not only sustaining near–real-time throughput
but doing so \emph{efficiently}. Both sites operate on comparatively
carbon-intensive grids based on the last twelve months of
ElectricityMaps~\cite{electricitymaps}:
\SI{0.672}{kg\,CO_2/kWh} in South Africa (SA) and \SI{0.321}{kg\,CO_2/kWh} in
Western Australia (WA). Continuous
\SIrange{1}{5}{MW} operation therefore emits 
\SIrange{5.9}{29.5}{kt\,CO_2/yr} (SA) and 
\SIrange{2.8}{14.1}{kt\,CO_2/yr} (WA), excluding embodied carbon.
\textbf{This makes computational efficiency a direct lever on SKA SDPs'
environmental footprint.}

Because operational emissions scale directly with compute efficiency, the limitations of current platforms become critical. In practice, for SKA-scale workloads, low arithmetic intensity and utilisation mean:
\begin{itemize}
  \item \textbf{Scalability failures:} many imaging workloads encounter strong-scaling limits in practice due to communication, synchronization, and serial overheads, yielding diminishing throughput gains as hardware is added.
    \item \textbf{Higher operational and capital costs:} more nodes are required to offset low utilisation, increasing electricity and cooling demands.
    \item \textbf{Higher carbon emissions:} fewer images, catalogues, and time series are delivered per ton of CO$_2$ emitted.
\end{itemize}

Energy efficiency, utilisation, and parallel scaling are therefore not just performance metrics but \textbf{climate performance indicators}, coupling algorithmic design and hardware co-design to measurable carbon reduction. A factor-of-two improvement in application-level GFLOP/s/W directly halves SDP operational emissions at fixed science throughput. Moreover, utilisation improvement reduces the total cost of ownership (TCO) by lowering both capital expenditure (CAPEX) and operational expenditure (OPEX)~\cite{alvarez2025ceo}.

\subsection{Gaps in Current Evaluation Practice}
\label{sec:codesign-motivation}

\begin{table}[htbp]
\scriptsize
\setlength{\tabcolsep}{3pt}
\renewcommand{\arraystretch}{0.9}
\centering
\caption{
Metric coverage across imaging tools: full pipelines (WSClean, DDFacet, the Bluebild Imaging++ pipeline BIPP, and the snapshot-based WS-Snapshot), the IDG algorithmic kernel, and the reduced-precision FPGA implementation FPGA-RP.
Symbols: explicit (\checkmark), implicit (\(\sim\)), missing (---).
Metrics are detailed in Table \ref{tab:metric-summary}. }
\label{tab:mega-gap-table}

\begin{tabular}{l c c c c c c c}
\toprule
\textbf{Metric Category} &
\textbf{WSClean} &
\textbf{IDG} &
\textbf{DDFacet} &
\textbf{FPGA} &
\textbf{BIPP} &
\textbf{WS-S.} &
\textbf{A.CAMP} \\
\midrule
\textbf{Pub.\ Year} &
2014 & 2018 & 2020/23 & 2022 & 2025 & 2025 & -- \\
\midrule

\multicolumn{8}{l}{\textbf{System-level (heterogeneous node / pipeline)}} \\
Time-to-solution $T_c$
  & \checkmark & \checkmark & \checkmark & \checkmark & \checkmark & \checkmark & \checkmark \\
Energy-to-solution $E_c$
  & ---        & \(\sim\)    & ---        & \checkmark & ---        & ---        & \checkmark \\
Throughput $\Theta$ (vis/s)
  & \checkmark & \checkmark & \(\sim\)    & \checkmark & \checkmark & \(\sim\)    & \checkmark \\
Energy efficiency $\eta_E$
  & ---        & \(\sim\)    & ---        & \checkmark & ---        & ---        & \checkmark \\

\midrule
\multicolumn{8}{l}{\textbf{Hardware platform-level  (CPU / GPU / FPGA / ASIC)}} \\
Utilisation / occupancy $U$
  & ---        & \(\sim\)    & ---        & \checkmark & ---        & ---        & \checkmark \\
Memory bandwidth $B_{\mathrm{mem}}$
  & ---        & \checkmark  & ---        & \checkmark & ---        & ---        & \checkmark \\

Memory efficiency $\eta_{mem}$
  & ---        & \(\sim\)    & ---        & --- & ---        & ---        & \checkmark \\

Peak memory usage $M_{\mathrm{peak}}$
  & \(\sim\)   & \(\sim\)     & \(\sim\)   & \(\sim\)   & \(\sim\)   & \checkmark & \checkmark \\

\midrule
\multicolumn{8}{l}{\textbf{Sustainability }} \\
Carbon-to-solution $C_c$
  & ---        & ---        & ---        & ---        & ---        & ---        & \checkmark \\
Carbon efficiency $\eta_C$
  & ---        & ---        & ---        & ---        & ---        & ---        & \checkmark \\

\midrule
\multicolumn{8}{l}{\textbf{Economics }} \\
Cost per job  
  & ---        & ---        & ---        & ---        & ---        & ---        & \checkmark \\

\bottomrule
\end{tabular}
\end{table}

Recent wide-field imaging studies have delivered significant advances in
algorithmic sophistication and numerical fidelity, yet their evaluation
methodology remains largely \emph{performance-centric}. As summarised in
Table~\ref{tab:mega-gap-table}, runtime and dirty-image RMS are consistently
reported across WSClean~\cite{offringa2014wsclean},
BIPP~\cite{tolley2025bipp}, WS-Snapshot~\cite{wu2025performance},
IDG~\cite{van2018image}, FPGA-RP~\cite{corda2022reduced}, and
DDFacet~\cite{tasse2023ddfacet,monnier2020parallelisation}, but key
architecture-level metrics—power, energy-to-solution, energy efficiency,
roofline characterisation, bandwidth sensitivity, and hardware utilisation—
are either missing or only implicitly analysed. Even in DDFacet, where
multi-node parallelisation has been demonstrated~\cite{monnier2020parallelisation},
no measurements of power or energy per job are reported. This
limits and slows quantitative comparison across CPUs, GPUs, FPGAs, and emerging
accelerators towards optimizing efficiency metrics.

System-level behaviour exhibits similar fragmentation. Pipeline-stage
breakdowns and parallel efficiency are available in DDFacet’s distributed
implementation~\cite{monnier2020parallelisation}, and partially in
IDG~\cite{van2018image} and FPGA-RP~\cite{corda2022reduced}, but metrics such
as device occupancy, data locality, bandwidth limits, and end-to-end dataflow
behaviour are rarely treated systematically. This hinders the identification of
system bottlenecks that directly influence scalability and energy efficiency in
SKA-scale deployments.

\textbf{None of the surveyed works reports
carbon-to-solution, carbon efficiency, or any economic metrics}, despite
their increasing relevance given SKA's strict power and environmental
constraints.
Overall, the literature reveals a fragmented evaluation landscape in which no
existing study jointly assesses architecture-level efficiency, system-level
behaviour, algorithmic fidelity, and sustainability. To date, only one imaging study has reported total energy-to-solution~\cite{corda2022reduced}, but only at the algorithmic kernel level (IDG), not for the entire imaging pipeline (WSClean). \textbf{astroCAMP} addresses
this gap by providing a unified, reproducible, and cross-layer metric framework
that enables principled algorithm--hardware co-design and supports
energy-, carbon-, and fidelity-aware optimisation for large-scale imaging systems.

\section{astroCAMP Framework}

\begin{figure}[!htbp]
    \centering
    \includegraphics[trim={0 0 0 1.5cm},clip,width=0.9\linewidth]{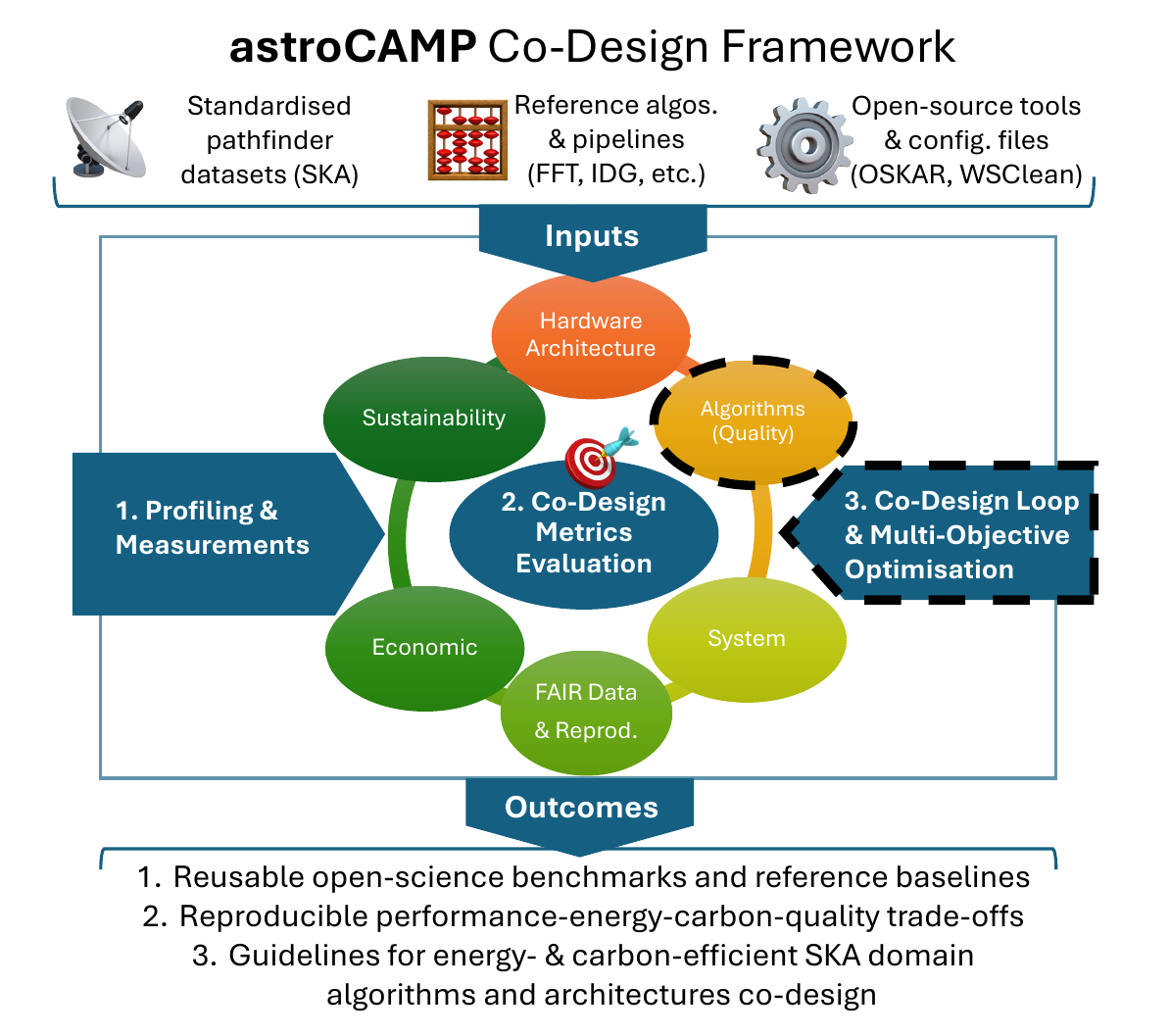}
    \caption{\textbf{astroCAMP} co-design framework.
    Dotted modules are intended for community extension.}
    \Description{Block diagram of the astroCAMP framework showing three stages: profiling and measurement, cross-layer metric evaluation, and a multi-objective co-design loop, with dotted modules marked as community-extensible.}
    \label{fig:astrocamp-overview}
\end{figure}

\textbf{astroCAMP} provides the missing infrastructure for reproducible,
cross-layer evaluation of radio-interferometric imaging software and hardware. 
It unifies algorithmic, architectural, and sustainability metrics into a single
benchmarking and co-design workflow (Fig.~\ref{fig:astrocamp-overview}), enabling
transparent comparison of imaging algorithms and hardware platforms.
The framework enables designers to examine:
\begin{enumerate}
    \item how algorithmic parameters (precision, kernel sizes, tiling) interact with CPU/GPU/FPGA platforms, supporting Outcome~1 by producing reusable, cross-platform performance and fidelity baselines;
    \item which configurations satisfy SKA~SDP power and throughput limits and populate Pareto-optimal performance--energy--quality trade spaces, delivering Outcome~2; and
    \item how close commodity hardware can approach SKA efficiency targets and where accelerators are justified, directly informing Outcome~3 on energy- and carbon-efficient co-design.
\end{enumerate}

\subsection{Benchmarking and Co-Design Methodology}

astroCAMP provides a reproducible methodology for evaluating radio-interferometric
imaging pipelines across heterogeneous architectures. The framework standardises
benchmark cases, in\-put–out\-put configurations, and a cross-layer metric suite
(performance, energy, carbon, utilisation, and scientific fidelity), enabling
fair comparison of algorithmic variants and hardware platforms under realistic
SKA-class constraints.
The proposed methodology ensures that results obtained on different systems and with
different imaging algorithms remain directly comparable. It consists of three
non-overlapping stages (illustrated in Fig.~\ref{fig:astrocamp-overview}):

\textbf{1. Profiling \& Measurements.}
Each benchmark is executed under controlled and repeatable conditions. We wait for node to reach a steady thermal and power
state. 
Rack-level power distribution units (PDUs) provide energy measurements to calculate
energy-to-solution~$E_c$, while platform telemetry (e.g., the Power Measurement Toolkit, PMT~\cite{corda2022pmt}, for CPUs, and the NVIDIA Management Library, NVML, for GPUs) is timestamp-aligned with PDU traces to validate device-level behaviour. From these measurements we derive time-to-solution~$T_c$, throughput, and
carbon-to-solution~$C_c$ using region-specific carbon intensity. All scripts,
configurations, and reproducibility guidelines are openly available~\cite{astroCAMP2025}.

\textbf{2. Co-Design Metrics Evaluation.}
For each benchmark configuration—including algorithmic parameters (e.g., IDG kernel
sizes) and platform choice (CPU, GPU, FPGA)—we assemble a consistent vector of
cross-layer metrics defined in Section~\ref{sec:codesign-metrics}. These metrics
capture performance, energy, cost, sustainability, and system constraints (e.g., SKA~SDP
node power caps). This stage performs \textit{metric computation only}: it normalises and
validates measurements across systems, preparing a unified metric vector that serves
as input to optimisation. In the current release, system and platform metrics are
evaluated using PREESM~\cite{PREESM2017}, and sustainability/economic metrics via
CEO-DC~\cite{alvarez2025ceo}. Algorithm-level quality metrics are intentionally
excluded in this release and form an open call for the community to converge on
standardised tolerances.

\textbf{3. Co-Design Loop \& Multi-Objective Optimisation.}
The metric vectors from Stage~2 feed into a structured design-space exploration (DSE) loop following the formulation in Section~\ref{sec:codesign-formulation}. The objective is to systematically evaluate alternative algorithm–ar\-chi\-tec\-ture mappings and solve the resulting multi-objective optimisation problem to produce reproducible Pareto frontiers and trade-off curves.
This stage provides the \textit{interpretation layer}: identifying efficient regions of the design space, exposing performance–energy\-qual\-i\-ty bottlenecks, and determining when approximate methods or domain-specific accelerators become necessary to meet SKA-scale efficiency targets.

\subsection{Co-design Metrics}
\label{sec:codesign-metrics}

\begin{table*}[t]
\scriptsize
\caption{\textbf{astroCAMP} co-design metrics. Core metrics (C) enter
the performance–energy–quality-cost optimisation; diagnostic metrics (D) support
interpretation and reproducibility.  
Type symbols: Direct = $\bullet$, Derived = $\circ$, Proxy/Model-based = $\triangle$.}
\label{tab:metric-summary}
\setlength{\tabcolsep}{2pt}
\renewcommand{\arraystretch}{1.08}

\begin{tabularx}{\textwidth}{@{}
  >{\centering\arraybackslash}m{0.025\textwidth}   
  >{\RaggedRight\arraybackslash}m{0.23\textwidth} 
  >{\centering\arraybackslash}m{0.04\textwidth}   
  >{\centering\arraybackslash}m{0.018\textwidth}   
  >{\centering\arraybackslash}m{0.022\textwidth}  
  >{\RaggedRight\arraybackslash}m{0.18\textwidth} 
  >{\RaggedRight\arraybackslash}X                 
@{}}
\toprule
\textbf{Layer} & \textbf{Metric / Formula} &
\textbf{Unit} & \textbf{Role} & \textbf{Type} &
\textbf{Instrumentation} & \textbf{Interpretation} \\
\midrule

\multirow{4}{*}{\rotatebox{90}{\textbf{System (Pipeline) }}}

& Time-to-solution $T_c$
& s & C & $\bullet$
& POSIX \texttt{time}, workflow logs, scheduler timestamps
& End-to-end wall-clock runtime of the workload on the full (possibly heterogeneous) system, where $T_c$ is the elapsed execution time. \\

& Energy-to-solution $E_c = \int_0^{T_c} P(t)\,dt$
& J & C & $\bullet$
& Rack/node PDUs, CPU RAPL, GPU NVML, PMT traces
& Total electrical energy consumed by the system, integrating instantaneous power $P(t)$ over runtime $T_c$. \\

& Throughput $\Theta = N/T_c$
& vis/s  & C & $\circ$
& MS/FITS logs, visibility/image counters
& System-level processing rate, where $N$ is the number of visibilities or images processed during runtime $T_c$. \\

& Energy efficiency $\eta_E = N/E_c$
& vis/J & C & $\circ$
& Derived from $N$ and $E_c$
& System-level science throughput per joule, using processed data volume $N$ and energy-to-solution $E_c$. \\

& Data locality  
$L_d = V_{\mathrm{local}}/V_{\mathrm{total}}$
& -- & D & $\triangle$
& Darshan I/O profiler, filesystem telemetry
& Fraction of total I/O volume served locally, where $V_{\mathrm{local}}$ and $V_{\mathrm{total}}$ are local and total I/O volumes. \\

\midrule

\multirow{3}{*}{\rotatebox{90}{\textbf{Hardware Platform}}}

& Utilisation $U = t_{\mathrm{active}}/t_{\mathrm{total}}$
& -- & D & $\bullet$
& \texttt{nvidia-smi}, ROCm tools, \texttt{perf}, Prometheus exporters
& Fraction of time a device (CPU socket, GPU, FPGA, etc.) is actively running kernels, where $t_{\mathrm{active}}$ is active compute time and $t_{\mathrm{total}}$ is total wall time. \\

& Memory bandwidth $B_{\mathrm{mem}} = \mathrm{Bytes}/T_c$
& GB/s & D & $\bullet$
& Hardware counters, Intel VTune, NVIDIA Nsight
& Sustained data-movement rate per platform, with $B_{\mathrm{mem}}$ indicating whether kernels are memory-bound. \\

& Memory efficiency $\eta_{\mathrm{mem}} = \mathrm{Bytes}/E_c$ 
& GB/J & D & $\bullet$
& Derived from memory-traffic counters and energy-to-solution $E_c$
& Useful data movement per joule at system level, computed as total DRAM bytes transferred divided by system energy-to-solution $E_c$. Complements $B_{\mathrm{mem}}$ by quantifying the energy cost of memory traffic. \\

& Peak memory usage $M_{\mathrm{peak}}$
& GB & D & $\bullet$
& \texttt{/proc/meminfo}, \texttt{nvidia-smi}, cgroups, container telemetry
& Maximum resident memory footprint observed on a platform during execution, constraining batch size and problem scaling. \\

\midrule

\multirow{6}{*}{\rotatebox{90}{\textbf{Algorithmic\ Quality}}}

& Dirty-image RMS  
$\sigma_{\mathrm{dirty}} = \sqrt{\frac{1}{N} \sum (I_i - \bar I)^2}$
& Jy/beam & C & $\bullet$
& CASA/WSClean \texttt{imstat}, PyBDSF
& Noise and artefact level in the dirty image, using pixel intensities $I_i$, mean value $\bar{I}$, and resulting RMS $\sigma_{\mathrm{dirty}}$. \\

& PSNR / SSIM,  
PSNR $= 10\log_{10}(I_{\max}/\mathrm{MSE})$
& dB / -- & C & $\bullet$
& scikit-image, OpenCV
& Fidelity of reconstruction $\hat{I}$ vs.\ reference $I_{\mathrm{ref}}$, using maximum pixel $I_{\max}$ and mean-squared error; SSIM measures structural similarity between $\hat{I}$ and $I_{\mathrm{ref}}$. \\

& Dynamic range  
$DR = I_{\max}/\sigma_{\mathrm{res}}$
& -- & C & $\bullet$
& Residual-image statistics
& Ratio of peak intensity $I_{\max}$ to residual RMS $\sigma_{\mathrm{res}}$, determining detectability of faint emission. \\

& Astrometric error  
$\epsilon_{\mathrm{astro}} = \frac{1}{N}\sum\|\mathbf{x}_i - \mathbf{x}_i^{\mathrm{ref}}\|$
& arcsec or px & C & $\bullet$
& PyBDSF catalogues
& Average positional discrepancy between reconstructed and reference sources, where $\mathbf{x}_i$ and $\mathbf{x}_i^{\mathrm{ref}}$ are measured and reference positions. \\

\midrule

\multirow{2}{*}{\rotatebox{90}{\textbf{Sustain.}}}

& Carbon-to-solution  
$C_c = E_c \, \kappa(t,r)$
& gCO$_2$e & C & $\triangle$
& Grid carbon-intensity APIs (e.g.\ electricityMap) + $E_c$
& Total carbon footprint of a run, using measured energy $E_c$ and grid carbon intensity $\kappa(t,r)$ at time $t$ and region $r$. \\

& Carbon efficiency  
$\eta_C = N/C_c$
& vis/gCO$_2$e & D & $\circ$
& Derived from $N$ and $C_c$
& Visibilities processed per gram CO$_2$ emitted, normalising science throughput $N$ by carbon cost $C_c$. \\

\midrule

\multirow{3}{*}{\rotatebox{90}{\textbf{Econom.}}}

& Total cost ownership  
$C_{\mathrm{TCO}} = C_{\mathrm{capex}} + C_{\mathrm{opex}}$
& € & D & $\bullet$
& Procurement \& operations records
& Lifetime system cost, combining capital cost $C_{\mathrm{capex}}$ and operational cost $C_{\mathrm{opex}}$. \\

& Cost per job  
$C_E = E_c\,p_E$
& € & C & $\circ$
& Electricity tariff + $E_c$
& Monetary cost per workload, using energy-to-solution $E_c$ and electricity price $p_E$. \\

& Cost efficiency  
$\Theta / C_{\mathrm{TCO}}$
& vis/€ & D & $\circ$
& Throughput + TTO data
& Science operations delivered per euro invested, using throughput $\Theta$ and total cost $C_{\mathrm{TCO}}$. \\

\bottomrule
\end{tabularx}
\end{table*}

astroCAMP includes a compact, cross-layer metric suite that links hardware
behaviour, energy use, economic cost, scientific throughput, and fidelity. Table~\ref{tab:metric-summary} summarises all metrics,
including their formulas, units, role, and instrumentation, and interpretation. 
The objective function is to minimize time-, energy-, carbon, and cost-to-solution while maximizing science throughput. Only the core metrics (C) are target for optimization, while diagnostic metrics  (D) support the interpretation of results by revealing system-level and platform-level bottlenecks—such as remote data access, uneven work distribution across CPUs and GPUs, and delays caused by slower devices. 

The framework
captures \textit{system-level (pipeline)} key performance indicators, including time-to-solution $T_c$,
en\-er\-gy-to-so\-lu\-tion $E_c$, and scientific throughput (visibilities/s or images/s), as well as \textit{hardware platform (architecture)}  metrics to inform co-design decisions, such memory bandwidth, memory efficiency, and
utilisation.
\textit{Algorithmic quality} is quantified via dirty-image root-mean-square (RMS), peak signal-to-noise ratio and structural similarity (PSNR/SSIM), dynamic range, and astrometric and photometric errors from catalogue comparisons,
capturing the scientific impact of approximations such as reduced precision,
coarse $w$-stacking, or non-uniform Fast Fourier Transform (NUFFT) kernel truncation. \textit{Sustainability} metrics pair
energy use with regional carbon intensity $\kappa(t,r)$ to compute
carbon-to-solution $C_c$, enabling joint optimisation of energy, carbon, and
image fidelity. \textit{Economic} metrics---including cost per job $C_E = E_c\,p_E$ and
amortised total cost of ownership $C_{\mathrm{TCO}}$---connect computational
choices to financial constraints relevant for large-scale SKA deployments.
Together, the full metric
suite supports transparent, multi-objective co-design across CPU, GPU, FPGA,
and ASIC platforms.


\section{Extensible Multi-Objective Co-Design Formulation}
\label{sec:codesign-formulation}

astroCAMP provides a \emph{general and extensible} co-design formulation that
unifies algorithmic, architectural, and system-level degrees of freedom.
Rather than prescribing specific algorithms or hardware, astroCAMP defines a
\emph{stable metric backbone} (Table~\ref{tab:metric-summary}) onto which
current and future imaging algorithms (e.g., WSClean, IDG, BIPP), accelerator
technologies (CPUs, GPUs, FPGAs, ASICs), and workflow runtime can be
instantiated.

\textbf{Design space.} Any imaging configuration is represented as a design point
$x = (a, h, s) \in \mathcal{X}$, shown in Fig.~\ref{fig:codesign-overview},
where:
\begin{itemize}
    \item $a$ are \textit{algorithmic parameters}: precision,
          convolution kernel size, NUFFT or $w$-stacking order, snapshot
          parameters, visibility tiling, task fusion.
    \item $h$ are \textit{hardware parameters}: processor type, accelerator count,
          memory hierarchy, power caps, thread/block geometry.
    \item $s$ are \textit{system/workflow parameters} for mapping $a$ to $h$: parallel decomposition,
          data-locality strategy, I/O staging, buffering and batching,
          workflow orchestration.
\end{itemize}

\begin{figure}[b]
\centering
\resizebox{\linewidth}{!}{%
\begin{tikzpicture}[
    font=\large,
    >=stealth,
    node distance=1cm,
    thick,
    transform shape=false
]

\node[anchor=west] at (-0.2,4.6) {\textbf{Design Space} $x = (a,h,s)$};

\draw[rounded corners, thick] (-0.5,0) rectangle (4,4);

\draw[->] (0,0) -- (3.5,0) node[midway,below] {Algorithmic $a$};
\draw[->] (0,0) -- (0,3.5) node[midway,left] {Hardware $h$};
\draw[->] (0,0) -- (-0.7,-0.7) node[pos=0.9,below left] {System $s$};

\foreach \x/\y in {0.8/0.6, 1.5/1.3, 2.2/0.9, 3/2.5, 1.1/3.2, 2.7/3.3, 0.5/2.2} {
  \fill[blue!50] (\x,\y) circle (2.2pt);
}

\node[align=center, anchor=north east] at (3.4,2.1)
    {All feasible \\ configurations};

\draw[->, thick] (4.2,2) -- (6.2,2) node[midway,above]{\small Metric evaluation};

\node[anchor=west] at (6.5,4.6)
  {\textbf{Objective Space} $(T_c,E_c,C_c, C_{TCO},\Delta Q)$};

\begin{scope}[shift={(7,0)}]

\draw[->] (0,0) -- (4,0)
    node[midway,below] {$T_c$};
\draw[->] (0,0) -- (0,4)
    node[midway,left] {$E_c$};

\foreach \x/\y in {
    3.2/2.8, 2.7/3.2, 3.5/1.5,
    1.5/3.4, 2.8/2.2, 1.8/2.7
} {
  \fill[gray!60] (\x,\y) circle (2.2pt);
}

\draw[red!70,very thick]
    (0.7,3.4) -- (1.1,2.4) -- (1.8,1.7) -- (2.5,1.2) -- (3.2,0.8);

\foreach \x/\y in {
    0.7/3.4, 1.1/2.4, 1.8/1.7, 2.5/1.2, 3.2/0.8
} {
  \fill[red!80] (\x,\y) circle (2.5pt);
}

\node[red!80, align=center] at (2.5,2.5)
    {\textbf{Pareto} \\ \textbf{frontier}};

\node[align=center] at (2,-0.8)
    {\small (Projection of 5D objective space)};

\end{scope}
\end{tikzpicture}
}%
\caption{%
Overview of the astroCAMP co-design formulation.  
}
\label{fig:codesign-overview}
\end{figure}

On the left side of Fig. \ref{fig:codesign-overview}, we see how each configuration $x$ is evaluated on a fixed workload $w$
(e.g.\ an SKA-Low or SKA-Mid benchmark) and is represented as a point in the design space of algorithmic, hardware platforms, and system parameters.

\textbf{Metrics}. For the DSE with astroCAMP framework, shown in Fig.~\ref{fig:codesign-overview}, we consider the following core metrics detailed in Table~\ref{tab:metric-summary}:
\[
T_c(x),\;
E_c(x),\;
C_c(x),\;
\Theta(x),\;
C_{TCO},\;
Q(x).
\]
Here, $T_c$ is time-to-solution, $E_c$ is Energy-to-solution,
$C_c$ is Carbon-to-solution, $\Theta$ is throughput
(e.g., visibilities/s or pixels/Joule), $C_{TCO}$ is the total cost of ownership, and $Q(x)$ is a
quality tuple (e.g., dirty-image RMS noise, PSNR).
A high-fidelity reference configuration defines $Q_{\mathrm{ref}}$.
We measure scalarised quality loss as:
\[
\Delta Q(x) = d\!\left(Q_{\mathrm{ref}},\, Q(x)\right),
\]
where $d$ is a weighted distance in quality space.

\textbf{Multi-objective optimisation problem}. We formalise co-design as a multi-objective optimisation problem subject to scientific and operational constraints.
\begin{equation}
\label{eq:moo-main}
    \min_{x \in \mathcal{X}}
    \quad
    \bigl(
        T_c(x),\;
        E_c(x),\;
        C_c(x),\;
        \Delta Q(x)
    \bigr)
\end{equation}

\paragraph{Scientific quality constraint:}
\[
\Delta Q(x) \le \Delta Q_{\max}.
\]

\paragraph{SDP operational constraints:}
\begin{align}
    P_{\mathrm{avg}}(x) &\le P_{\max}
        &&\text{(site power cap, e.g.\ 2 MW)}, \\
    \Theta(x) &\ge \Theta_{\min}
        &&\text{(minimum throughput / survey cadence)}, \\
    x &\in \mathcal{X}_{\mathrm{valid}}
        &&\text{(resource bounds, feasibility)}.
\end{align}

Constraints couple algorithmic, hardware, and workflow choices to architectural, sustainability, and economic layers through energy, power, and carbon metrics. The formulation is modular: new algorithmic parameters extend $a$, new hardware platforms extend $h$, and new workflow or runtime options extend $s$. Additional metrics can be added to the objective or constraints without changing the optimisation structure. In this way, astroCAMP defines a \emph{stable design space schema} that still accommodates future algorithmic and architectural innovation.

Because the objectives in Eq.~\eqref{eq:moo-main} conflict, \textit{no single optimal
configuration exists}. We therefore compute the \textit{Pareto frontier}
\[
\mathcal{P}
    = \Bigl\{
        x \in \mathcal{X}_{\mathrm{valid}}
        \;\Big|\;
        x \text{ is not dominated across }
        (T_c, E_c, C_c, \Delta Q)
      \Bigr\},
\]
where ``not dominated’’ means that there is no
$x' \in \mathcal{X}_{\mathrm{valid}}$ such that
$
T_c(x') \le T_c(x), E_c(x') \le E_c(x), 
C_c(x') \le C_c(x),
\Delta Q(x') \le \Delta Q(x),
$
with at least one inequality strict. 

Each $x \in \mathcal{P}$ represents a distinct, scientifically valid, and
operationally feasible trade-off (e.g., ``fast but power-hungry'', ``high
precision but slower'', ``energy-efficient but low throughput''). The construction of $\mathcal{P}$ is done via structured parameter sweeps at the software and/or the hardware architecture level.  Evaluation of co-design metrics can be performed using simulation tools~\cite{sarkar_lightningsim_2023}, model-based design~\cite{PREESM2017} or surrogate models. 
Automated design space exploration \cite{zaourar2023deca} can rely on Bayesian optimization \cite{fu2025high}, or evolutionary search to analyze how both new (BIPP, WS-Snapshot) and
legacy algorithms (IDG) populate the multi-metric
design space.

\section{astroCAMP Benchmark Suite and Datasets}

The benchmaks and datasets are designed to be large enough to stress memory, I/O, and communication layers while remaining manageable for design-space
exploration experiments. 
We provide open, standardised benchmark cases, representative of radio-interferometric imaging workloads.  
Each case includes:
\begin{itemize}
    \item Synthetic and pathfinder datasets derived from realistic telescope configurations (e.g.\ SKA-Low), generated with the Oxford SKA simulator \texttt{OSKAR};
    \item Reference outputs from validated CPU implementations (e.g.\ the WSClean wide-field imaging pipeline and its IDG kernel);
    \item Configuration files specifying baseline distribution, frequency range, integration time, and field of view (FoV).
\end{itemize}

The datasets were generated using \texttt{OSKAR}~\cite{OSKAR} (version OSKAR-2.11.2-dev 2025-07-17 ddb65ed) using the SKA-Low 512-station full configuration. Sixteen datasets were produced by varying the number of timesteps (time samples) over 1, 8, 64, 128, and 256, and the number of channels over 1, 8, 64, 128, and 256. Each timestep is integrated over 10 seconds. The phase centre was set to 25.0 and -30.0 degrees in right ascension and declination, respectively. The start frequency was 151\,MHz with increments of 1\,MHz. Sources were drawn from the GaLactic and Extragalactic All-sky MWA (GLEAM) catalogue~\cite{gleam2017} to cover a 40-degree FoV. 
Data volumes range from megabytes to terabytes, enabling tests from workstation to cluster scale. 
Each dataset includes:
\begin{itemize}
    \item Raw power and timing logs at the kernel and pipeline level, together with energy metrics.
    \item Ground-truth dirty image to compare algorithmic quality metrics.
    \item Configuration templates for CPU and GPU runs.
\end{itemize}

\section{Benchmark Evaluation and Results}

This section benchmarks SKA-Low workloads and datasets to demonstrate the practical impact of our co-design methodology. We report results from a subset of metrics (performance, energy efficiency, sustainability, and economy), then analyse heterogeneous WSClean+IDG execution, CPU-only strong scaling, location-dependent carbon/cost efficiency for identical workloads, and PREESM-based design-space exploration for performance–energy–utilization trade-offs toward Pareto-optimal SKA-scale operating regions.

\subsection{Experimental Setup}

Benchmarks were executed on a Lenovo \textit{ThinkSystem SR675 V3} node of the Green500 \textit{KUMA} cluster \cite{kuma_node_spec}, with dual AMD EPYC~9334 CPUs, 371~GB RAM, 6.4~TB NVMe, and four NVIDIA H100 GPUs (94~GB). All runs used exclusive node access and allocated 16~CPU cores, one H100 GPU, and one quarter of RAM (92.2~GB; Slurm limit 5900~MB/core).

\begin{figure}[!htbp]
  \centering
  \includegraphics[trim={0 0 0 2.2cm},clip,width=0.9\linewidth]{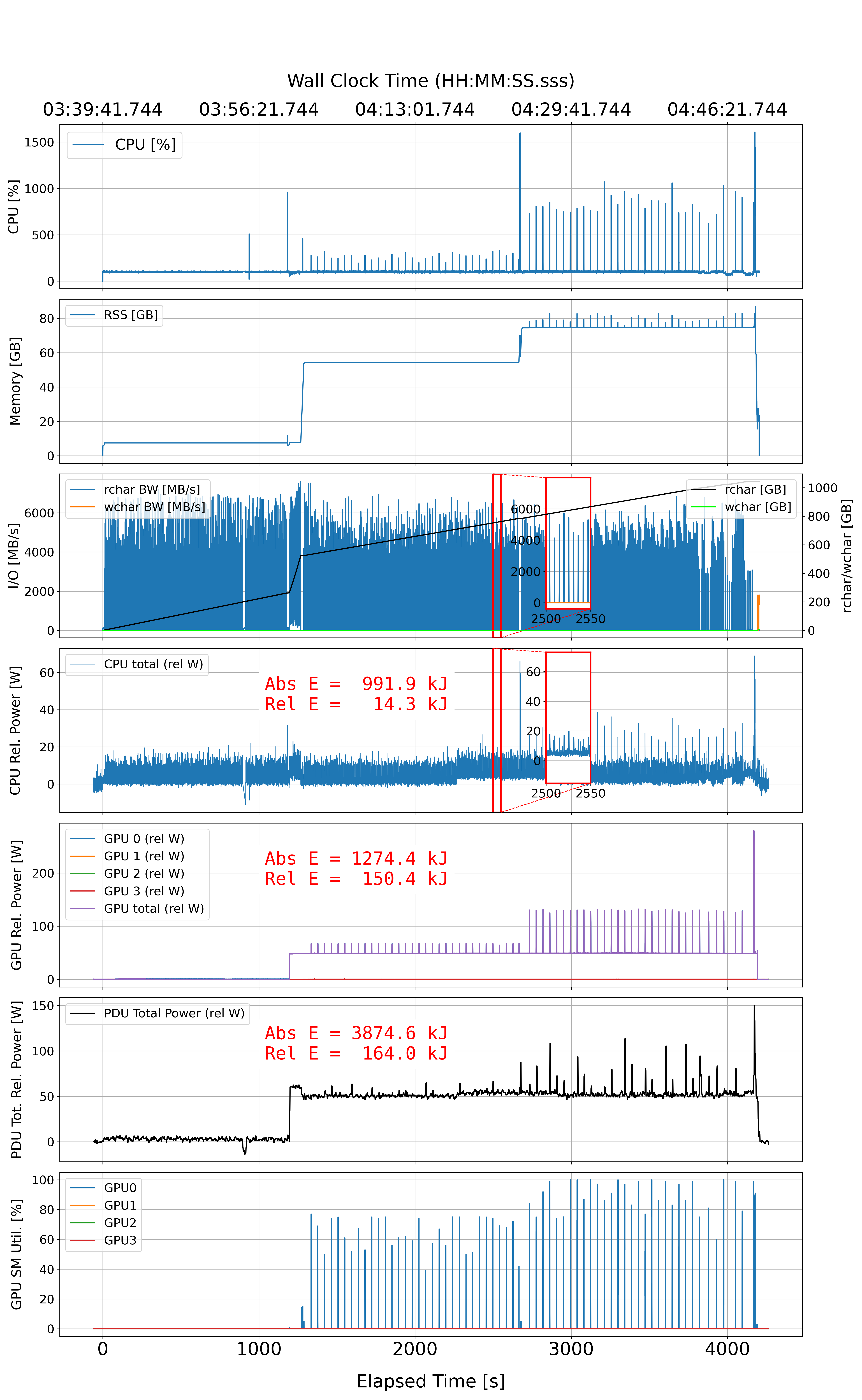}
  \caption{Profiling of WSClean + IDG on GPU: 256 timesteps, 256 channels, full SKA-Low, for a 32k $\times$ 32k pixels image.}
  \Description{Time-series traces of CPU utilization, memory footprint, I/O activity, and CPU/GPU/system power deltas during a WSClean+IDG run, showing bursty multi-core activity over a single saturated CPU thread and intermittent GPU power draw.}
  \label{fig:slurm}
\end{figure}

The server cost was obtained by configuring a system with comparable specifications,
resulting in \SI{29875.39}{USD}~\cite{lenovo_sr675_v3}.
Embodied hardware emissions were estimated by combining the emissions reported by
NVIDIA for the H100 GPU (\SI{176}{kgCO$_2$e}~\cite{nvidia_h100_pcf}) with the
emissions of the remaining system components modeled using Boavizta’s Datavizta
tool~\cite{boavizta_datavizta}, yielding a total embodied emission of
\SI{513}{kgCO$_2$e}. The latter assumes a Genoa-class CPU architecture, Samsung
memory modules, and a Micron SSD.
We assume that the CAPEX is amortized over a six-year lifetime.

The data were processed with \texttt{WSClean}~\cite{offringa2014wsclean} in \texttt{IDG}~\cite{vandertol2018idg} GPU mode. For each of 16 datasets, we fixed FoV and varied image size ($4096^2$, $8192^2$, $16\,384^2$, $32\,768^2$; pixel scales 17.578, 8.789, 4.394, 2.197 arcsec). For each of the $25 \times 4 = 100$ runs, we executed: warmup, 120\,s pause, monitored run, 120\,s pause. Warmup compiles CUDA kernels; the monitored run reuses compiled kernels. The pre-run pause allows hardware to settle.
Power measurements were recorded using node-level PDUs, which provided a 1\,s power average sampled every 5\,s. The PDU measurements were cross-validated with software monitors (RAPL, on the CPU side, and NVML on the GPU side), using the PMT library wrapper~\cite{corda2022pmt}. Fig.~\ref{fig:slurm} exemplifies CPU utilisation, memory footprint, I/O intensity, and CPU/GPU/system power (as deltas from idle). CPU activity is mostly confined to one core (100\,\%) with short multi-core bursts, indicating poor CPU-side scaling. GPU power remains well below the H100 thermal design power (TDP) and streaming-multiprocessor (SM) utilisation appears in short bursts, indicating that work supply is intermittent and/or constrained by memory-access effects.

\subsection{Heterogeneous WSClean+IDG Deployment}

\begin{figure}[!htbp]
  \centering
  \includegraphics[width=0.9\linewidth]{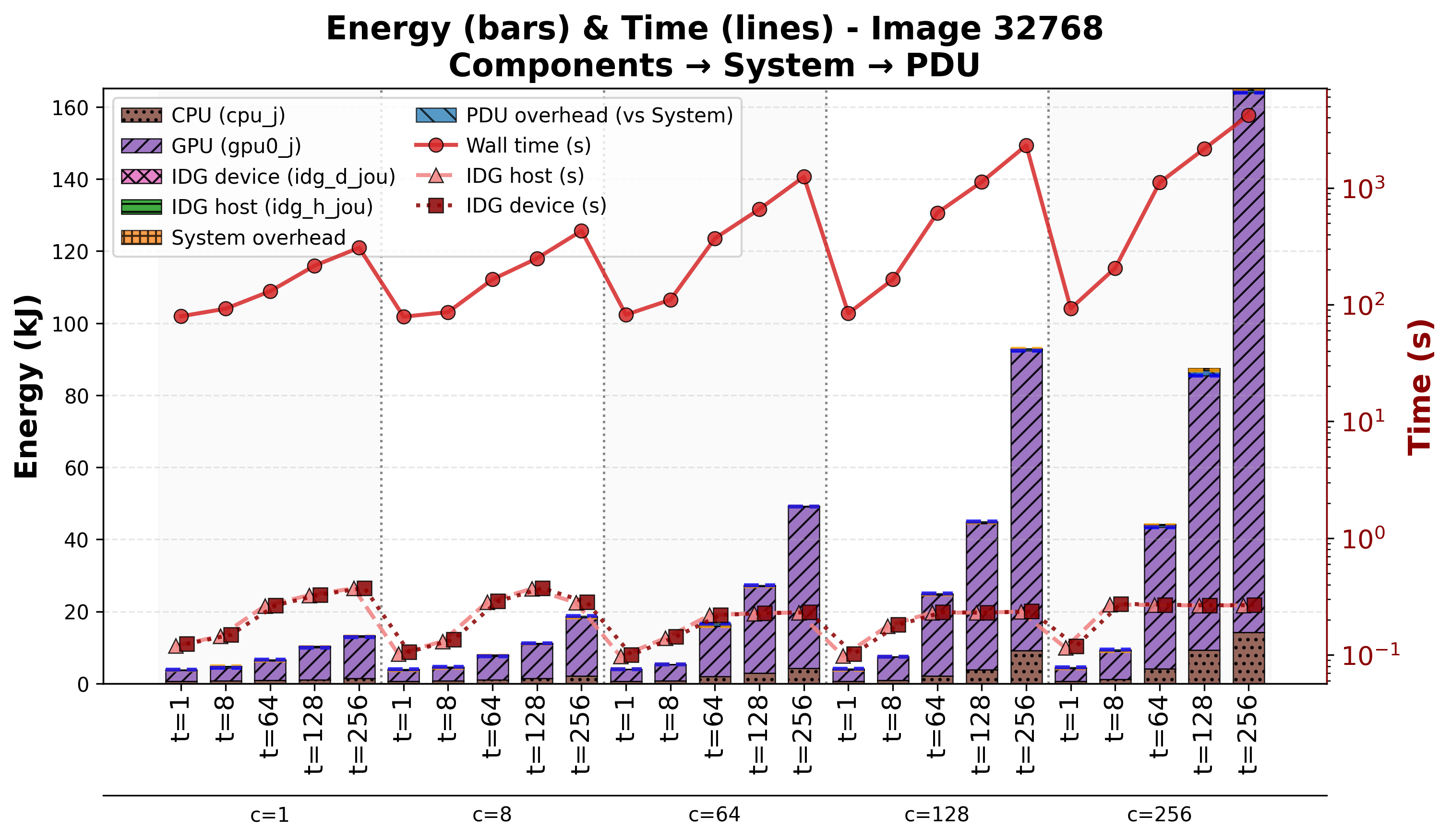}
  \caption{Energy hierarchy at $32{,}768^2$ across all $(t,c)$ configurations. Stacked bars (left axis, kJ, hatch-coded): CPU, GPU, IDG device, IDG host, system overhead, PDU overhead. Red lines on the log-scale right axis: wall time, IDG host time, IDG device time.}
  \Description{Stacked bar chart of per-component energy at $32768^2$ image size across timestep and channel configurations, with hatch-coded layers for CPU, GPU, IDG device, IDG host, system overhead, and PDU overhead, plus three red time-series lines on a log-scale right axis showing wall time, IDG host time, and IDG device time.}
  \label{fig:energy-hierarchy}
\end{figure}

WSClean+IDG is a heterogeneous pipeline: the CPU performs batching, metadata preparation, and scheduling, while the GPU executes IDG gridding and interpolation. Throughput is high only when the CPU feeds work fast enough to keep the GPU busy.
Fig.~\ref{fig:energy-hierarchy} shows a clear hierarchy for $32768^2$ across $(n_{\mathrm{times}}, n_{\mathrm{chans}})$: GPU and IDG-device layers dominate the active-compute share (about 60--90\%), CPU and IDG-host layers stay nearly flat, and the system- and PDU-overhead layers add a sizable rack-level component above the node-level measurement. GPU energy rises from about 5--10\,kJ to about 165\,kJ at $t256$–$c256$, whereas CPU/host-side IDG energy remains nearly constant, confirming accelerator-driven scaling. The right-axis red lines (wall time, IDG host time, IDG device time, on a log scale) grow together with energy over nearly four orders of magnitude, and channel-heavy workloads deliver similar throughput with lower wall time and total energy than time-heavy cases.

\begin{figure*}[!t]
  \centering
  \includegraphics[trim={0 0 0 3cm},clip,width=\linewidth]{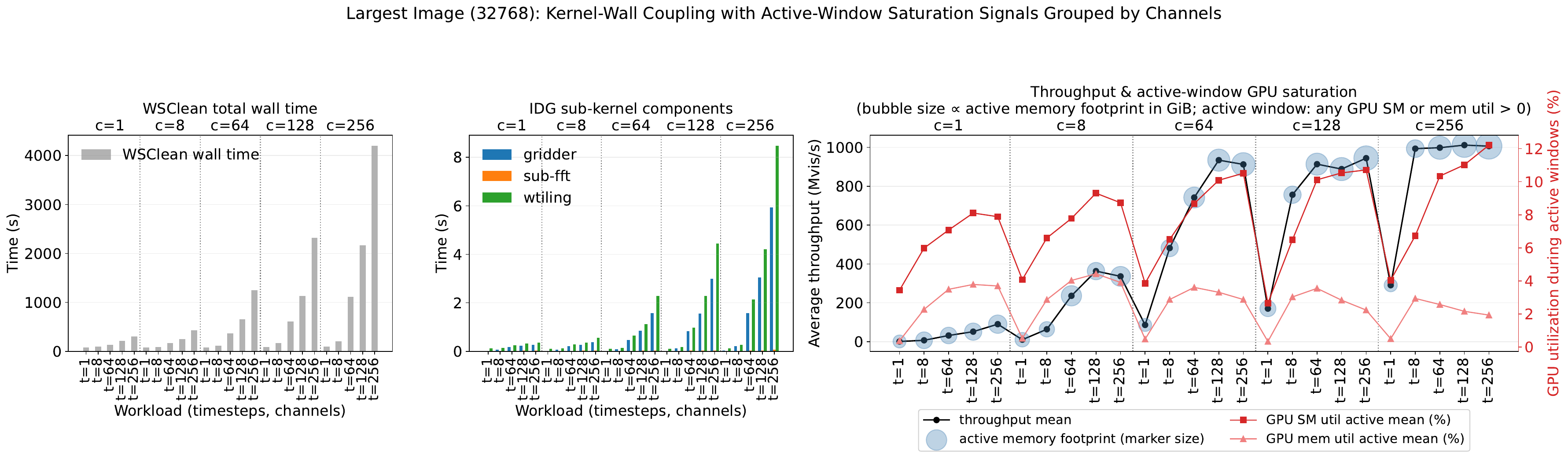}
    \caption{Kernel--wall coupling and active-window GPU saturation at $32{,}768^2$. \emph{Left:} \textsc{WSClean} wall time. \emph{Middle:} IDG sub-kernels (gridder, sub-FFT, $w$-tiling) stacked, on a much smaller time scale. \emph{Right:} throughput (left axis, Mvis/s, black; bubble size $\propto$ active memory footprint) with GPU SM and memory utilisation in active windows on the right axis.}
  \Description{Three-panel figure for the 32768-squared workload, grouped by channel count. Left panel: bar chart of total WSClean wall time per (t,c). Middle panel: stacked bar chart of IDG sub-kernel times (gridder, sub-FFT, w-tiling). Right panel: throughput in Mvis/s with bubbles sized by active-window memory footprint, plus two right-axis lines for active-window GPU SM utilisation and GPU memory utilisation in percent.}
  \label{fig:idg-saturation-signals}
\end{figure*}

\begin{figure}[!htbp]
  \centering
  \includegraphics[trim={1cm 0.1cm 1cm 2cm},clip,width=0.93\linewidth]{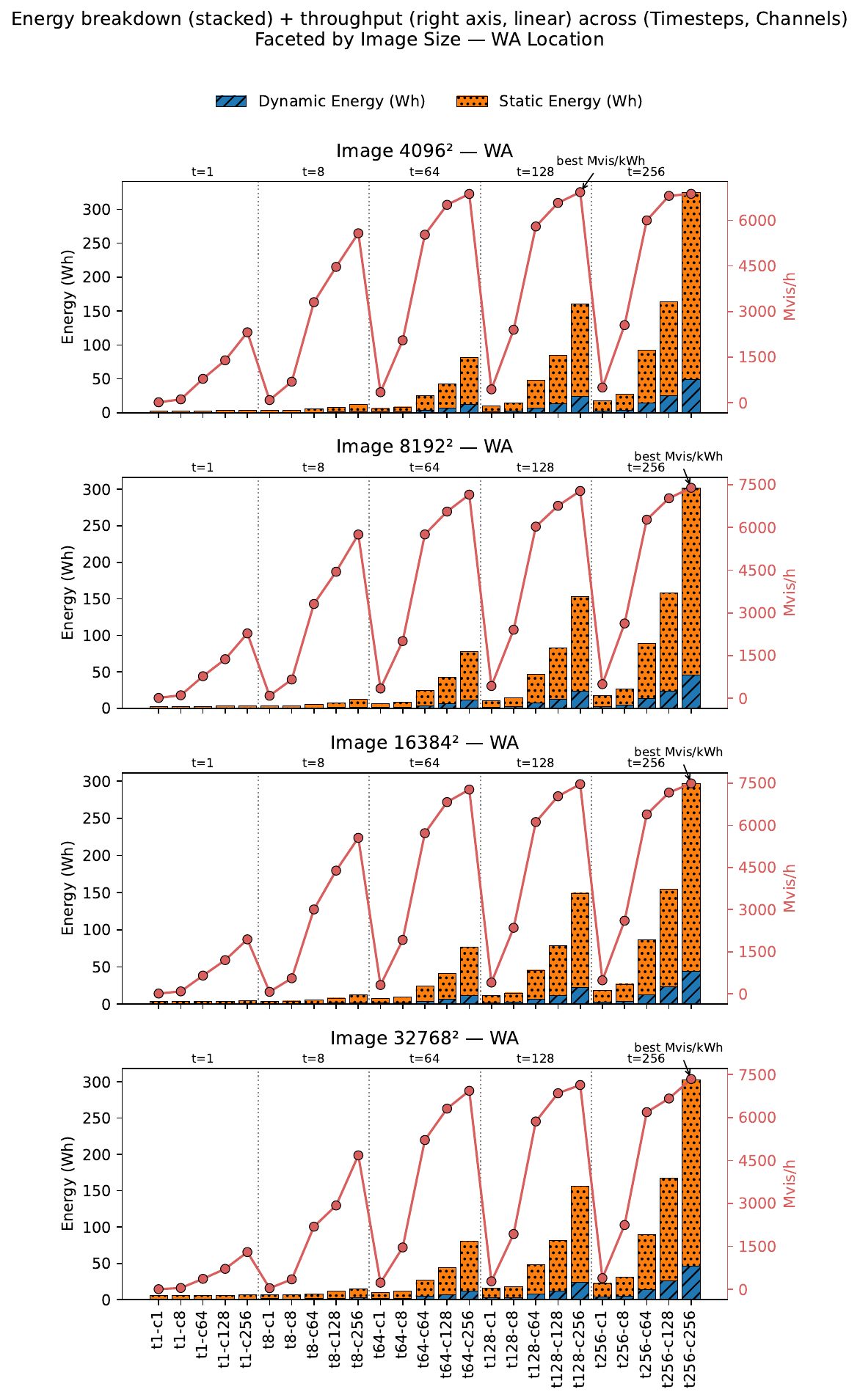}
  \caption{WSClean+IDG energy and throughput on SKA-Low (WA), faceted by image size. Stacked bars: dynamic (blue hatched) + static (orange dotted) energy in Wh (left). Red line: throughput in Mvis/h (right). Annotated marker in each panel: best Mvis/kWh.}
  \Description{Four-panel facet plot for SKA-Low at WA showing, per image size, stacked dynamic and static energy bars on the left axis and an overlaid throughput line in Mvis per hour on the right axis, across all combinations of timesteps and channels, with the best Mvis-per-kWh point annotated in each panel.}
  \label{fig:energy-time-carbon}
\end{figure}

\begin{figure}[!htbp]
  \centering
  \includegraphics[width=0.9\linewidth]{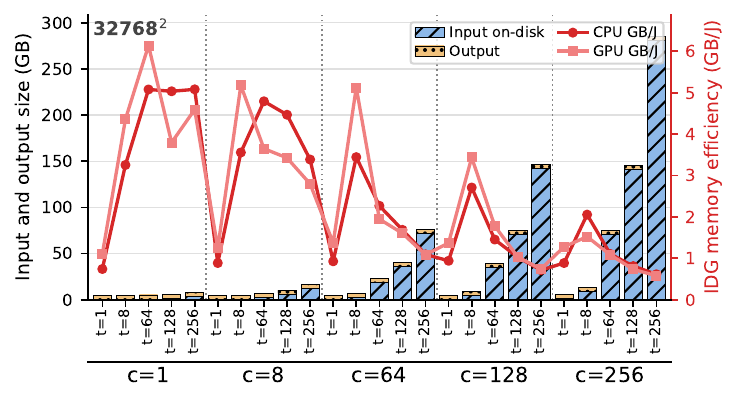}
  \caption{Payload footprint and IDG memory efficiency at $32{,}768^2$. Stacked bars (left, GB): input on-disk (blue hatched) and output (orange dotted). Lines (right, GB/J): CPU (dark-red circles) and GPU (pink squares) IDG execution.}
  \Description{Single-panel figure at 32768 squared with stacked bars on the left axis showing on-disk input and output volume in GB across timestep and channel configurations grouped by channels, plus two right-axis lines reporting IDG memory efficiency in GB per Joule for CPU and GPU execution.}
  \label{fig:payload-footprint}
\end{figure}

We decompose total energy into \emph{static} and \emph{dynamic} terms. Static energy is the idle baseline (CPUs, GPUs, memory, storage, networking, cooling, PSU losses), estimated as one quarter of PDU idle power times runtime (because we use a single GPU node of the four available, a fourth of CPU and memory); dynamic energy is the incremental compute energy from integrating GPU power plus one quarter of CPU power over runtime. Static energy remains dominant across configurations (about \SI{80}{\percent}–\SI{85}{\percent}), indicating persistent hardware under-utilisation even at the largest workloads.

To understand \emph{why} utilisation stays so low, we first examine kernel--wall coupling. Fig.~\ref{fig:idg-saturation-signals} shows that increasing $c$ raises explicit IDG kernel work, but end-to-end WSClean wall time remains much larger than the summed sub-kernel times, so application-level gains are not proportional. Throughput flattens before active-window SM utilisation saturates and while active memory utilisation stays modest, indicating a workflow-level bottleneck. This matches Nsight Compute: some kernels are efficient while active, but those states are not sustained across the full workflow timeline, so gains are more likely from batching, overlap, synchronisation reduction, and staged-data reuse than from kernel-only tuning.

This kernel--wall gap manifests directly in the energy/throughput trade-off. Fig.~\ref{fig:energy-time-carbon} shows how energy and throughput vary across the astroCAMP dataset configurations: as visibility dimensions increase, throughput improves initially but rapidly saturates, while wall time and total energy grow super-linearly. The trend reflects increasing data-movement and orchestration overheads for larger working sets rather than a change in the fundamental computational bottleneck. Therefore, naïvely scaling imaging parameters increases operational cost and carbon footprint without delivering commensurate scientific throughput. This observation is underscoring the need for efficiency-aware configuration selection for SKA-scale deployments.

Fig.~\ref{fig:payload-footprint} complements this view with payload-footprint and memory-efficiency trends at $32768^2$: increasing channels mainly grows the input payload, whereas the output term remains fixed per image size (0.0625, 0.25, 1.0, and 4.0~GiB for $4096^2$ to $32768^2$). The GB/J curves indicate that larger spectral workloads do not translate into proportional memory-efficiency gains; the device path is generally above host, but both are sublinear versus footprint growth, confirming that scaling is increasingly input-dominated at larger $(t,c)$. The co-design implication is to prioritise channel-aware batching, fewer redundant visibility passes, lower orchestration overhead, and scheduling that keeps efficient GPU phases dominant for longer.

\begin{figure}[!htbp]
  \centering
  \includegraphics[width=\linewidth]{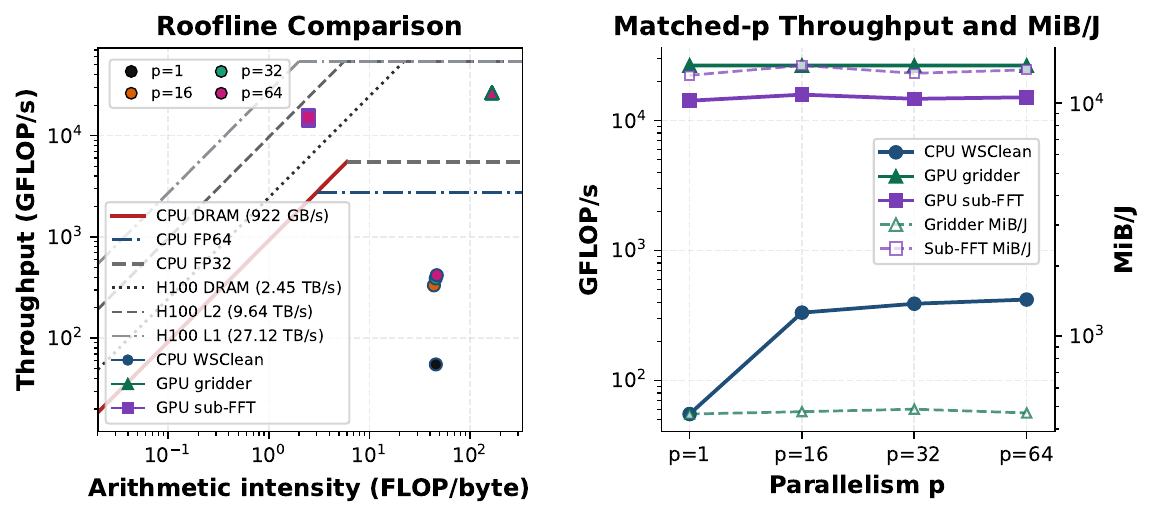}
  \caption{GPU-vs-CPU roofline view for WSClean stacking ($16{,}384^2$, $t=c=256$). \emph{Left:} CPU and H100 ceilings; marker shape = workload (CPU WSClean $\bullet$, GPU gridder $\blacktriangle$, GPU sub-FFT $\blacksquare$), fill colour = parallelism $p$. \emph{Right:} matched-$p$ throughput (GFLOP/s, left) and energy efficiency (MiB/J, right, dashed).}
  \Description{Two-panel roofline figure for WSClean stacking at t=256, c=256, 16384 squared. Left panel: log-log roofline with CPU and H100 ceilings and three marker shapes (circle, triangle, square) for CPU WSClean, GPU gridder, GPU sub-FFT, color-coded by parallelism p in 1, 16, 32, 64. Right panel: matched-p plot showing throughput in GFLOP/s on the left axis and MiB per Joule on the right axis as dashed lines, across p=1,16,32,64.}
  \label{fig:roofline-cpu-gpu-stack}
\end{figure}

\begin{figure}[!htbp]
  \centering
  \includegraphics[width=\linewidth]{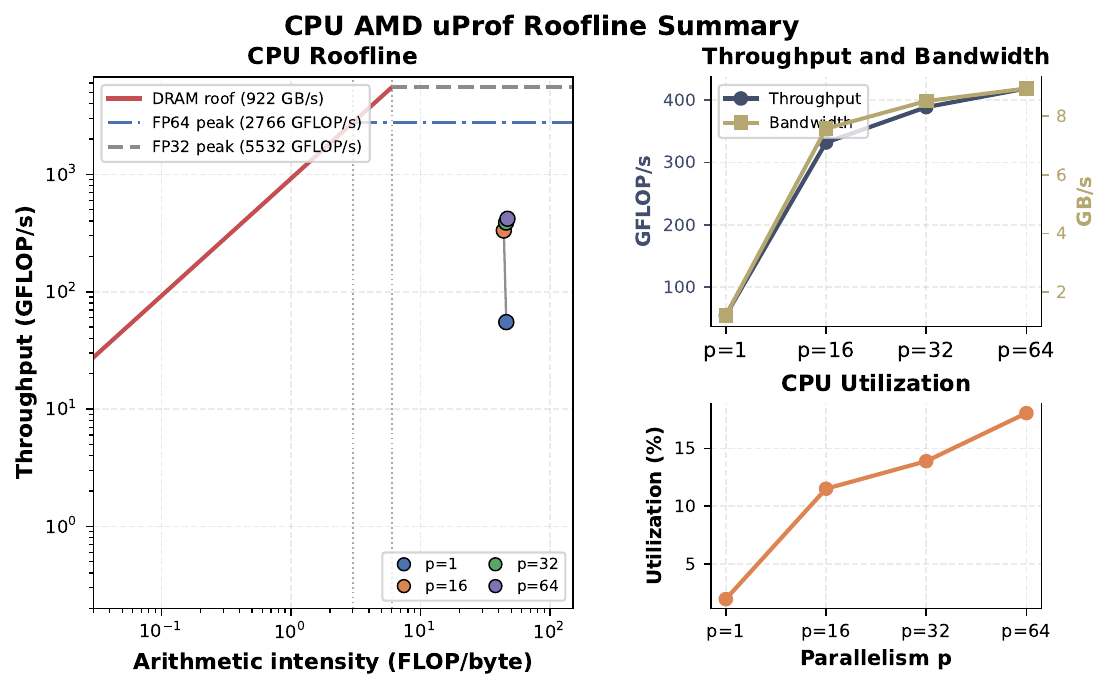}
  \caption{CPU AMD~uProf roofline summary for CPU-only WSClean stacking ($16{,}384^2$, $t=c=256$). \emph{Left:} roofline with DRAM, FP64 and FP32 ceilings; markers coloured by $p$. \emph{Top right:} throughput (GFLOP/s) and bandwidth (GB/s) vs.\ $p$. \emph{Bottom right:} CPU utilisation vs.\ $p$.}
  \Description{Three-panel CPU-only roofline summary at t=256, c=256, 16384 squared. Left: log-log roofline with DRAM, FP64 and FP32 ceilings and four operating points coloured by parallelism. Top right: throughput in GFLOP/s and bandwidth in GB/s versus parallelism. Bottom right: CPU utilisation in percent versus parallelism.}
  \label{fig:roofline-cpu-only-stack}
\end{figure}

A roofline study adds model-guided context. The left panel of Fig.~\ref{fig:roofline-cpu-gpu-stack} places CPU WSClean (circles), GPU gridder (triangles), and GPU sub-FFT (squares) against CPU and H100 ceilings: the GPU kernels operate near the H100 L1/L2 region only in their active windows, while the CPU operating points sit far below the FP64 ceiling despite arithmetic intensity above the DRAM ridge. The right panel reports the same workloads at matched~$p$, confirming that GPU GFLOP/s and energy efficiency (MiB/J) stay roughly flat with parallelism, whereas CPU GFLOP/s plateaus an order of magnitude lower. Fig.~\ref{fig:roofline-cpu-only-stack} zooms in on the CPU path: throughput and DRAM bandwidth grow sublinearly with $p$, and CPU utilisation rises only from $\sim2\%$ at $p=1$ to $\sim18\%$ at $p=64$. Together, the two views indicate that the end-to-end limits are better explained by orchestration, synchronization, and host-device pipeline effects than by a single DRAM-roof constraint.

\subsection{Strong CPU-only Scaling Limitations}

\begin{figure}[!htbp]
  \centering
  \includegraphics[width=0.6\linewidth]{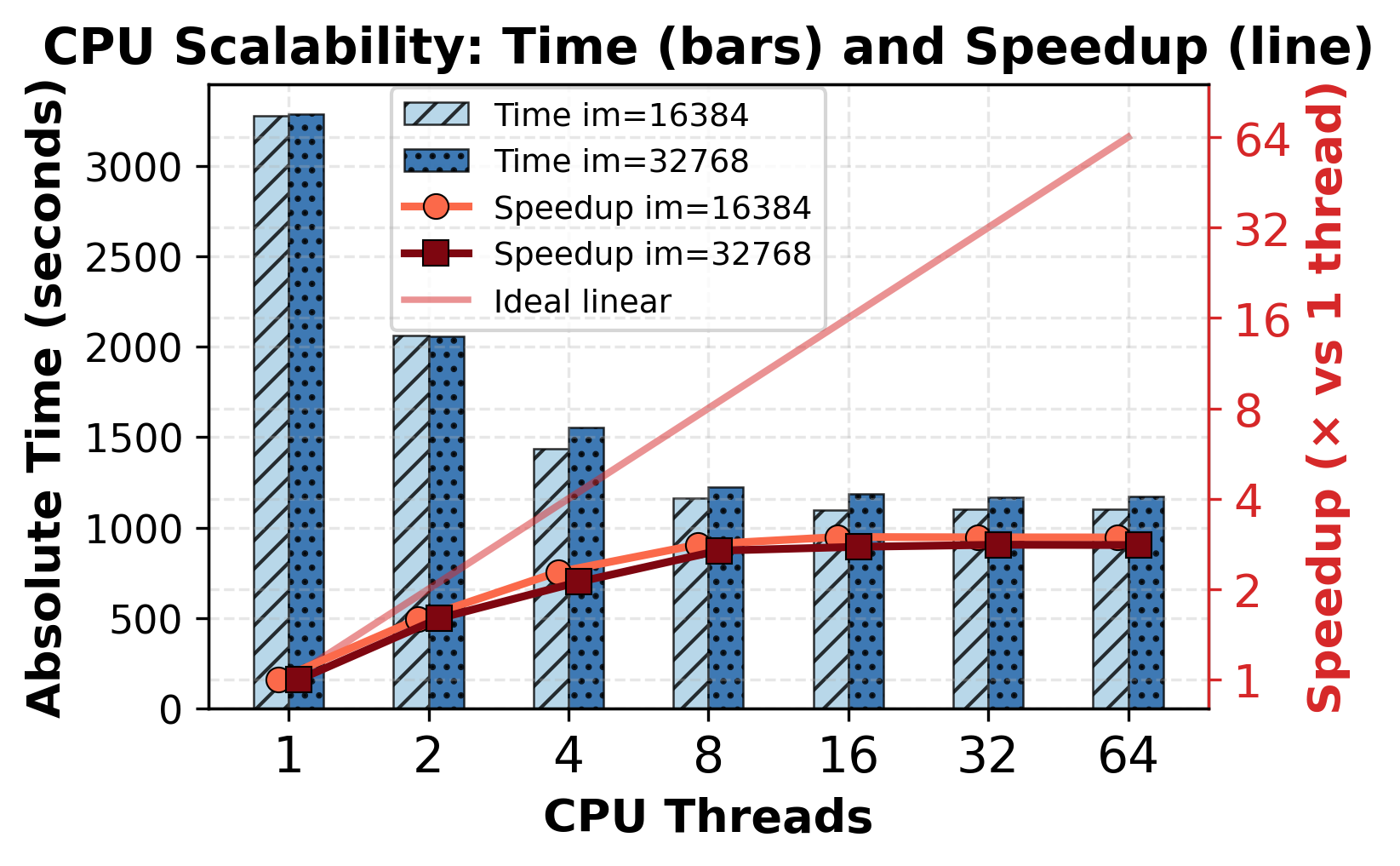}
  \caption{WSClean+IDG strong scaling on 1--64 CPU threads for two image sizes ($16{,}384^2$ hatched, $32{,}768^2$ dotted). Bars: wall time (left). Red lines: speedup vs.\ 1 thread (right); diagonal red line marks ideal scaling.}
  \Description{Bar-and-line chart of WSClean+IDG runtime and measured speedup from 1 to 64 CPU threads, with two image sizes (16384 and 32768) shown as hatched and dotted bar pairs and two corresponding speedup curves, against a diagonal ideal-linear reference line.}
  \label{fig:cpu-scaling}
\end{figure}

WSClean+IDG shows strong-scaling limitations across 1--64 CPU threads, as can be seen in Fig.~\ref{fig:cpu-scaling}. Consistent with~\cite{offringa2014wsclean}, performance improves from 1 to a few threads and then quickly stalls: beyond 16 threads, added cores yield essentially no further wall-time gain. End-to-end speedup saturates at \textbf{2.82$\times$ for $p\!\geq\!16$}, which corresponds to only \textbf{17.6\%} of ideal at $p=16$ and \textbf{4.4\%} at $p=64$, indicating that current stacks do not exploit many-core CPU servers efficiently for SKA-sized workloads.

Table~\ref{tab:wsclean_idg_cpu_gpu_wall_times} confirms this behavior on the flagship case: CPU-only runtime improves from one to 16 cores, then nearly plateaus, while GPU-enabled wall time remains nearly constant across CPU core counts because the dominant parallel work is already offloaded.

\begin{table}[htbp]
    \centering
    \caption{Execution times for \(t=256\), \(c=256\), \(16384^2\) with IDG on CPU or GPU across CPU core counts.}
    \label{tab:wsclean_idg_cpu_gpu_wall_times}
  \scriptsize
  \renewcommand{\arraystretch}{1.05}
  \setlength{\tabcolsep}{3pt}
  \begin{tabular}{lcccc}
        \toprule
    \textbf{Solution} & \textbf{1c} & \textbf{16c} & \textbf{32c} & \textbf{64c} \\
        \midrule
        WSClean + IDG CPU & 3:13:10 & 1:08:28 & 1:08:10 & 1:08:06 \\
        WSClean + IDG GPU & 1:09:27 & 1:09:57 & 1:09:14 & 1:08:45 \\
        \bottomrule
    \end{tabular}
\end{table}

\begin{figure}[!htbp]
  \centering
  \includegraphics[trim={4cm 0 2cm 2.8cm},clip,width=0.9\linewidth]{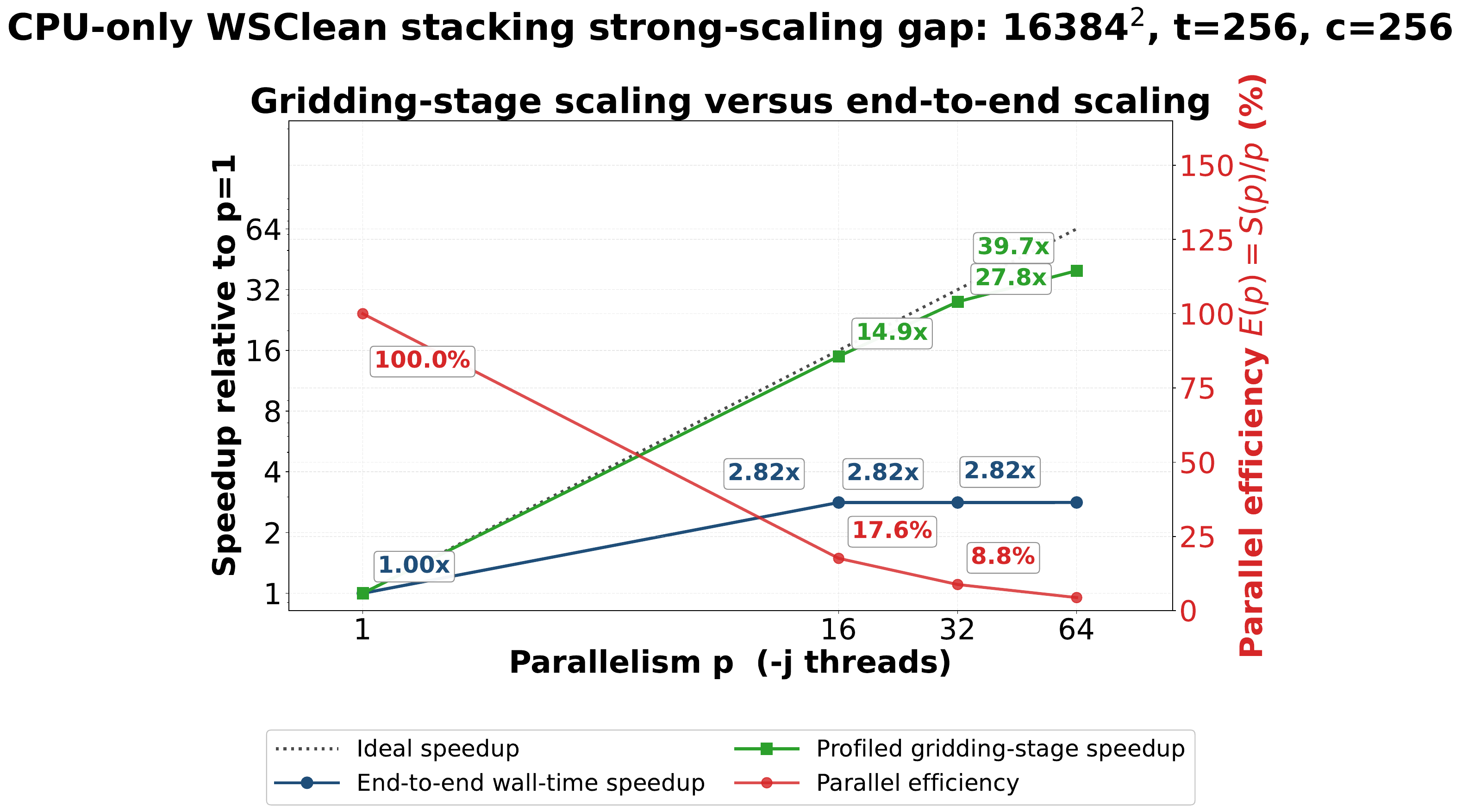}
  \caption{CPU-only WSClean stacking strong-scaling gap ($16{,}384^2$, $t=c=256$): end-to-end wall-time speedup (blue) vs.\ profiled gridding-stage speedup (green), with parallel efficiency on the right axis (red). Dashed: ideal speedup.}
  \Description{Strong-scaling plot for CPU-only WSClean stacking at 16384 squared, t=256, c=256: end-to-end wall-time speedup flattens at 2.82x while the gridding-stage speedup follows the ideal line up to about 39.7x at 64 threads, and parallel efficiency falls from 100\% to 4.4\%.}
  \label{fig:cpu-scaling-gap}
\end{figure}

Although bandwidth-dominated kernels can strong-scale in principle, here the losses are dominated by serial regions and orchestration overheads in the full application path. Fig.~\ref{fig:cpu-scaling-gap} makes this collapse explicit: while the profiled gridding stage alone scales near-ideally to $39.7\times$ at $p=64$, end-to-end wall time saturates at $2.82\times$, and overall parallel efficiency falls from $100\%$ at $p=1$ to $17.6\%$ at $p=16$ and $4.4\%$ at $p=64$. Flamegraph traces expose \texttt{libcasa\_tables} and \texttt{libcasa\_casa} (the CASA radio-astronomy I/O libraries used by WSClean) in effectively single-threaded sections, creating a serial metadata/I/O lane that absorbs added cores; moving from version~2 to version~4 of the Measurement Set format (MSv2 $\rightarrow$ MSv4) and to MSv4-capable I/O frameworks should ease this serialisation pressure. Consistently, I/O accounts for only $\sim$1\% of wall time but consists of many short, frequently misaligned requests: in the largest case we observe 99.66\% misaligned file requests and 280{,}024 reads below 1~MB (262{,}144 to the same data file), indicating that request aggregation and alignment remain relevant even when peak filesystem bandwidth is not saturated.

\subsection{Location-Dependent Efficiency for SDPs}

\begin{figure}[!htbp]
  \centering
  \includegraphics[width=0.45\textwidth]{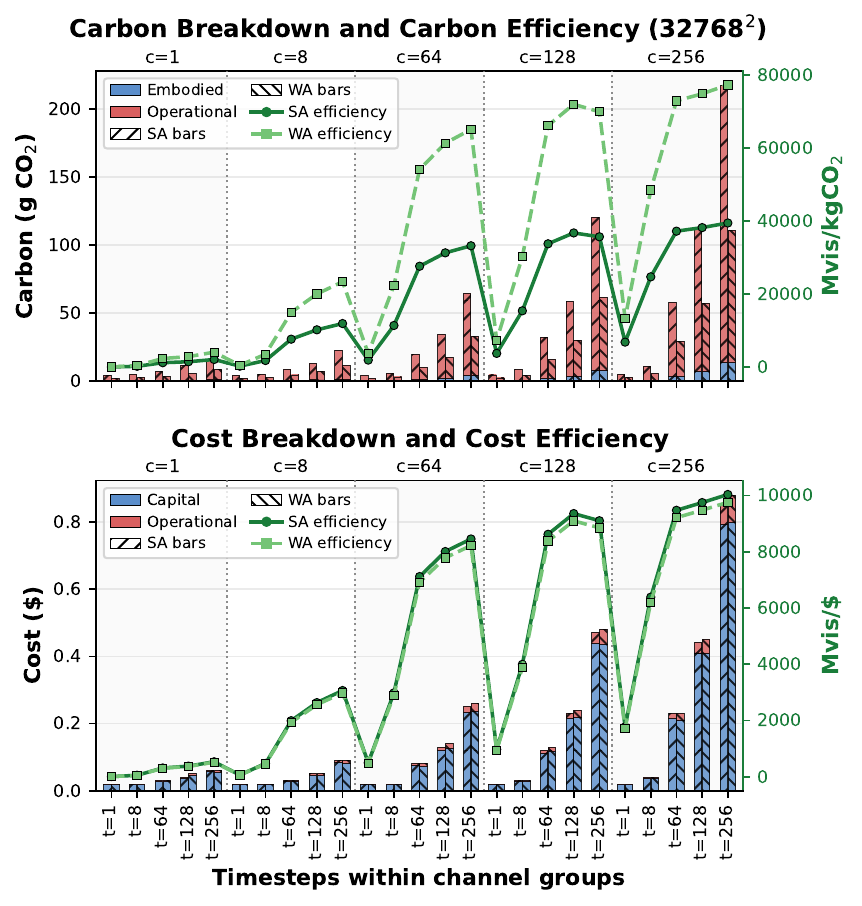}
  \caption{Carbon (top) and cost (bottom) breakdowns at $32{,}768^2$ across all $(t,c)$. Per $(t,c)$, paired stacked bars compare SKA-Mid/SA and SKA-Low/WA. Top stacks: embodied (blue) + operational (red); right axis: Mvis/kgCO\textsubscript{2}. Bottom stacks: capital (blue) + operational (red); right axis: Mvis/\$. SA: dark-green solid; WA: light-green dashed.}
  \Description{Two-panel figure for image size 32768 squared. Top panel: stacked carbon bars decomposed into embodied and operational components with two adjacent bars per timestep-channel combination representing SA and WA sites, plus right-axis lines for carbon efficiency in Mvis per kilogram CO2 (SA solid, WA dashed). Bottom panel: analogous stacked cost bars decomposed into capital and operational components, with right-axis lines for cost efficiency in Mvis per dollar.}
  \label{fig:carbon-cost-breakdown}
\end{figure}

In Fig.~\ref{fig:carbon-cost-breakdown}, we compare carbon and cost for the largest image size at the SKA SDP locations. Throughput is identical, but efficiency diverges with local grid intensity and electricity price. Operational emissions dominate carbon (90-95\%); WA attains higher carbon efficiency (Mvis/kgCO\textsubscript{2}) because grid intensity is lower (0.321 vs.\ 0.672~kg~CO\textsubscript{2}/kWh). Cost efficiency is CAPEX-dominated (89-92\%): WA's higher electricity price (0.27 vs.\ 0.19~\$/kWh in SA) yields lower Mvis/\$. Increasing timesteps/channels amortizes embodied carbon and CAPEX, so larger batches improve efficiency.

\subsection{Design Space Exploration with PREESM}

PREESM~\cite{PREESM2017} is a prime example of a DSE framework to study its extendability.
The diversity of metrics, specialized for a scientific field, calls for automated exploration support.
The current flow supports many \textit{system} and \textit{hardware platform}-level metrics, and more can be computed from those metrics based on user needs, using an existing model for fast evaluation.

Taking a step back to look at metric implementations,
\textit{algorithmic} metrics require the complete execution of the pipeline,
with potentially long execution, incompatible with DSE.
Future flows should model parameter influence on expensive metrics to avoid systematic evaluation~\cite{honorat:hal-03752645}. For example, changing quantization may require full execution, while changing hardware frequency may not.


We evaluate some of these metrics with PREESM by implementing the simulator, a subset of the DDFacet radio-astronomy pipeline that inverts an artificial sky image into visibilities. We run our designs on a KRIA KR260 system-on-module (SoM) board with a 4-core, 1.6\,GHz ARM Cortex-A53 processor and an Ultrascale+ FPGA clocked at 150\,MHz. We measure each design's resource usage, latency, and energy-to-completion at execution. The energy is obtained by measuring the base consumption of the device at idle and subtracting it from the consumption under load. Fig.~\ref{preesm_pareto_Ec} summarises our results. We define the occupancy metric as the weighted average of the percentage of CPU cores used and the utilisation of three FPGA resource families---configurable logic blocks (CLBs), digital signal processors (DSPs), and block RAMs (BRAMs): $avg(\textrm{CPU},\,avg(\textrm{CLB},\textrm{DSP},\textrm{BRAM}))$. This metric aims to measure the proportion of total available computing resources used. The FPGA-mappable computation is the 2-D FFT, split into two 1-D FFTs. The application is run on a 128x128 input image, resulting in a 128x128-point 2-D FFT, and outputs $\approx$98k visibilities.

\begin{figure}[!htb]
\centering

\begin{minipage}{0.45\linewidth}
\centering
\begin{tikzpicture}
  \begin{axis}[
    width=4.2cm, height=3.8cm,
    grid=both,
    grid style={line width=.1pt, draw=gray!10},
    major grid style={line width=.2pt,draw=gray!30},
    xlabel={latency (s)},
    ylabel={$E_c$ (J)},
    ylabel style={yshift=0pt},
    tick align=outside,
    enlargelimits=0.05,
    xmin=6.4, xmax=7,
    ymin=1, ymax=1.48,
  ]
    \addplot[ 
      only marks,
      mark=square*,
      mark size=2pt,
      draw=red!60!black,
      fill=red!60!white,
    ] coordinates {
    (6.89, 1.39)
    (6.77, 1.42)
    (6.89, 1.38)
    (6.48, 1.23)
    (6.54, 1.25)
    (6.5, 1.3)
    (6.46, 1.1)
    };
    \addplot[ 
      only marks,
      mark=*,
      mark size=2pt,
      draw=blue!60!black,
      fill=blue!60!white,
    ] coordinates {
    (6.54, 1.05)
    (6.54, 1.11)
    (6.54, 1.05)
    (6.57, 1.05)
    (6.53, 1.18)
    };
    \addplot[ 
    semithick,
    black,
    ] coordinates {
    (6.46, 1.47)
    (6.46, 1.1)
    (6.54, 1.05)
    (6.57, 1.05)
    (7, 1.05)
    };
  \end{axis}
\end{tikzpicture}
\end{minipage}%
\hfill
\begin{minipage}{0.45\linewidth}
\centering
\begin{tikzpicture}
  \begin{axis}[
    width=4.2cm, height=3.8cm,
    grid=both,
    grid style={line width=.1pt, draw=gray!10},
    major grid style={line width=.2pt,draw=gray!30},
    xlabel={latency (s)},
    ylabel={Occupancy (\%)},
    ylabel style={yshift=0pt},
    tick align=outside,
    enlargelimits=0.05,
    xmin=6.4, xmax=7,
    ymin=10, ymax=70
  ]
    \addplot[ 
      only marks,
      mark=square*,
      mark size=2pt,
      draw=red!60!black,
      fill=red!60!white,
    ] coordinates {
    (6.89, 31.2)
    (6.77, 22.6)
    (6.89, 22.5)
    (6.48, 43.7)
    (6.54, 35.1)
    (6.5, 35.1)
    (6.46, 68.7)
    };
    \addplot[ 
      only marks,
      mark=*,
      mark size=2pt,
      draw=blue!60!black,
      fill=blue!60!white,
    ] coordinates {
    (6.54, 12.5)
    (6.54, 25)
    (6.54, 50)
    (6.57, 12.5)
    (6.53, 25)
    };
    \addplot[ 
    semithick,
    black,
    ] coordinates {
    (6.46, 68.7)
    (6.48, 43.7)
    (6.5, 35.1)
    (6.54, 12.5)
    (7, 12.5)
    };
  \end{axis}
\end{tikzpicture}
\end{minipage}
\caption{PREESM Pareto fronts: latency vs.\ $E_c$ (left) and latency vs.\ occupancy (right). Square markers: CPU--FPGA; round: CPU-only; black line: non-dominated points.}
\Description{Two scatter plots produced by PREESM design-space exploration: latency versus energy (left) and latency versus occupancy (right), with square markers for CPU-FPGA designs and circular markers for CPU-only designs and a connecting Pareto-front line.}
\label{preesm_pareto_Ec}
\end{figure}

Using PREESM, we can explore the design space and derive the Pareto fronts shown in Fig.~\ref{preesm_pareto_Ec}, which projects each candidate mapping into two objective planes: latency vs.\ energy-to-completion (left) and latency vs.\ resource occupancy (right). This enables informed design decisions—such as the degree of parallelism and the mapping of tasks to CPU or FPGA—by explicitly trading off competing objectives.

\section{Discussion of Benchmark Evaluation}

The benchmarking results support a clear result-to-decision chain for SKA imaging co-design. Our evaluation of WSClean+IDG yields two evidence-based conclusions:
\begin{enumerate}
  \item \textbf{Pipeline-level scaling limits dominate:} CPU strong scaling collapses after low thread counts (Figs.~\ref{fig:cpu-scaling}, \ref{fig:cpu-scaling-gap}), CPU activity is bursty and mostly single-core in critical stages (Fig.~\ref{fig:slurm}), and end-to-end wall time remains far above summed kernel time even when active-window GPU utilization is nonzero (Fig.~\ref{fig:idg-saturation-signals}). Together with the roofline comparison (Fig.~\ref{fig:roofline-cpu-gpu-stack}), this points to orchestration/synchronization and serial-path effects rather than a single kernel-bound limit.
  \item \textbf{Efficiency gains are configuration- and site-dependent:} energy hierarchy and throughput trends (Figs.~\ref{fig:energy-hierarchy}, \ref{fig:energy-time-carbon}) show accelerator-dominated scaling with persistent static-energy overheads; payload trends indicate input-dominated growth at large $(t,c)$ (Fig.~\ref{fig:payload-footprint}); and carbon/cost efficiency diverges by location despite equal throughput (Fig.~\ref{fig:carbon-cost-breakdown}).
\end{enumerate}

Takeaway: use astroCAMP in three steps: (i) detect scaling/utilization bottlenecks (Figs.~\ref{fig:slurm}, \ref{fig:cpu-scaling}, \ref{fig:cpu-scaling-gap}, \ref{fig:idg-saturation-signals}), (ii) compare energy/carbon/cost efficiency across $(t,c)$ and sites (Figs.~\ref{fig:energy-time-carbon}, \ref{fig:carbon-cost-breakdown}, \ref{fig:payload-footprint}), and (iii) select the smallest configuration that meets science-quality tolerances.

\subsection{Why We Need HW--SW Co-Design?}
Optimizing SKA SDP requires joint SW and HW decisions because neither layer alone can optimize energy, carbon, cost, and runtime. The measured behavior in Figs.~\ref{fig:slurm}, \ref{fig:cpu-scaling-gap}, and \ref{fig:idg-saturation-signals} shows low sustained utilization and serial orchestration paths, with average utilization below \SI{5}{\percent}. As a result, static energy dominates total energy. Under this profile, improving SW efficiency and reducing runtime (with similar dynamic draw) can reduce total energy by up to \SI{81}{\percent}, with associated reductions of \SI{97}{\percent} in carbon emissions and \SI{32}{\percent} in total cost under WA assumptions.

The HW impact depends on SW-side utilization recovery. In the current profile, dynamic energy is about \SI{15}{\percent} of total energy, so HW-only efficiency tuning has limited system impact; under improved utilization, dynamic energy can rise toward \SI{80}{\percent}, making architecture and mapping choices substantially more influential. In that regime, dynamic-energy reductions map directly to operational savings that account for approximately \SI{78}{\percent} of total carbon emissions and \SI{25}{\percent} of total costs. Therefore, SW-only optimization is constrained by poor utilization, and HW-only optimization is constrained when dynamic energy is a small share; practical gains require co-optimizing both layers.

\subsection{Open Challenge for the SKA Community }

\textbf{Why is this challenge needed?} SKA documents define high-level goals---e.g.\ \emph{1\,\% astrometry} and spectral dynamic ranges of $10^{5}$--$10^{4}$---but not survey-level tolerances for flux accuracy, completeness, point-spread-function (PSF) residuals, polarisation purity, or spectral-line fidelity. Without these thresholds, the co-design community cannot determine which approximations (e.g., reduced precision, coarse $w$-stacking) are acceptable. With SKA power and cost envelopes fixed years ahead, the key question is which algorithm/hardware combinations maximize performance and energy efficiency without violating fidelity. This requires cross-layer metrics linking runtime, energy, carbon, utilization, and image quality, plus programme-level tolerances that are still missing~\cite{bacon2020cosmology}. 

\begin{tcolorbox}[
    colback=gray!5,
    colframe=black!40,
    boxrule=0.4pt,
    title=\textbf{How Much Quality can we Trade off for Efficiency?}
]
\emph{For each SKA Key Science Programme---Cosmic Dawn / Epoch of Reionisation (EoR), Galaxy Evolution and Cosmology, Cosmic Magnetism, and Time-Domain Astrophysics---define quantitative \textbf{application-level quality metrics and tolerances}.}
Examples include peak signal-to-noise ratio (PSNR) and structural similarity (SSIM), flux-scale accuracy, astrometric precision, polarisation purity, rotation-measure (RM) recovery error, spectral-line fidelity, time-series accuracy, and transient completeness.
Tolerances must be precise enough to guide hardware--software co-design
(precision choices, approximations, accelerators) while ensuring validity for:
\begin{itemize}
    \item 21\,cm EoR power spectra and tomographic cubes,
    \item all-sky continuum and deep extragalactic surveys,
    \item RM-grid measurements of cosmic magnetism,
    \item high-cadence transient and pulsar-timing searches.
\end{itemize}
\end{tcolorbox}

\textbf{Community engagement plan}.
astroCAMP follows an open-source, maintainer-guided model, hosted as a public repository~\cite{astroCAMP2025} where datasets, metrics, and reference implementations are curated by maintainers and extended via standard contribution workflows. This mirrors established HPC software practices, supporting transparent validation and gradual community convergence on shared quality metrics and tolerances.

\section{Conclusions}

In this paper, we have presented astroCAMP as an open benchmark and co-design framework for SKA imaging that couples representative datasets with unified runtime, energy, carbon, and cost metrics. The reported results show: (i) end-to-end limits are dominated by orchestration/synchronization and serial-path effects, not only kernel-level ceilings; (ii) CPU-only scaling is poor for SKA-sized workloads; and (iii) sustainability/economic outcomes depend on workload configuration and site context even at equal throughput. These findings translate into a practical strategy: prioritize pipeline utilization and data movement efficiency, then optimize architecture and scheduling under explicit quality tolerances.
We conclude with a community call to:
(1) establish science-driven imaging-quality tolerances,
(2) expand benchmark datasets,
(3) develop reference submissions for current and emerging imaging pipelines,
and
(4) build an open submission and validation process for the SKA
era.

\begin{acks}
This research was funded, in part, by the Agence Nationale de la Recherche grant agreements ANR-23-CE46-0010 and ANR-22-EXNU-0004, and the Swiss National Science Foundation grant no.~200021E\_220194: ``Sustainable and Energy Aware Methods for SKA (SEAMS)''. A CC~BY license is applied to the Author Accepted Manuscript (AAM) resulting from this submission, in accordance with the open-access conditions of the grant.
The authors thank the EPFL EcoCloud center, in particular Dr.~Xavier Ouvrard, and the team of IT Infrastructure and Operations of EPFL, in particular Junior Mbuyi, for providing access to its infrastructure for monitoring energy consumption in servers; and SCITAS in particular Daniel Filipe Jana, Yves Lopes and Nicolas Litchinko for setting up and providing access to a dedicated test node.
\end{acks}

\printbibliography
\end{document}